\documentclass[11pt]{article}
\pdfoutput=1
\usepackage{hyperref}
\usepackage{amsmath}
\usepackage{amssymb}
\usepackage{subcaption}
\usepackage{cite}
\usepackage{color,colordvi}
\usepackage{enumitem}

\usepackage{amssymb}
\usepackage{amsmath}
\usepackage{graphicx}
\usepackage{caption}
\usepackage{longtable}
\usepackage{verbatim,lipsum}
\usepackage{amsfonts}
\usepackage{hyperref}
\usepackage[dvipsnames]{xcolor}
\definecolor{rosso}{rgb}{0.92, 0.1, 0.05}
\definecolor{arancione}{rgb}{1., 0.5, 0}
\definecolor{verde}{rgb}{0, 0.35, 0.1}
\usepackage[normalem]{ulem}

\definecolor{Gray}{gray}{0.4}

\newcommand{\arXiv}[2]{\href{http://arxiv.org/pdf/#1}{{\tt [#2/#1]}}}
\newcommand{\arXivold}[1]{\href{http://arxiv.org/pdf/#1}{{\tt [#1]}}}

\def\d{{\rm d}}
\def\hbar{\overline{h}}

\newcommand{\be}{\begin{equation}}
\newcommand{\ee}{\end{equation}}
\newcommand{\bea}{\begin{eqnarray}}
\newcommand{\eea}{\end{eqnarray}}
\newcommand{\beq}{\begin{equation}}
\newcommand{\eeq}{\end{equation}}
\def\beqa{\begin{eqnarray}}
\def\eeqa{\end{eqnarray}}

\def\lsim{%
  \mathrel{\rlap{\lower4pt\hbox{\hskip0.5pt$\sim$}}\raise1pt\hbox{$<$}}%
  }  %less than or approx. symbol
\def\gsim{%
  \mathrel{\rlap{\lower4pt\hbox{\hskip0.5pt$\sim$}}\raise1pt\hbox{$>$}}%
  }  %greater than or approx. symbol

\def\d{\rm d}

\def\d{{\rm d}}

\def\MSbar{\relax\ifmmode\overline{\rm MS}\else{$\overline{\rm MS}${ }}\fi}

\newcommand{\eq}{\begin{equation}}
\newcommand{\eqx}{\end{equation}}
\newcommand{\eqn}{\begin{eqnarray}}
\newcommand{\bi}{\begin{itemize}}
\newcommand{\eqnx}{\end{eqnarray}}
\newcommand{\ei}{\end{itemize}}

\newcounter{hran}

\def\simlt{\stackrel{<}{{}_\sim}}
\def\simgt{\stackrel{>}{{}_\sim}}

\begin{document}
\begin{titlepage}
\begin{center}

\hfill CERN-TH-2018-233
\vskip 0.4 cm

\vspace{3cm}

{\LARGE\bf Implications of the detection of primordial \bigskip \\ gravitational waves for the Standard Model}
%\vspace{2cm}

\vspace{0.2cm}

\end{center}
\vskip 1cm

\begin{center}

{\bf G. Franciolini}$^{a}$, {\bf G.F.  Giudice}$^{b}$, {\bf D. Racco}$^{a, c}$ and  {\bf A. Riotto}$^{a}$
%\vspace{.6truecm}

\vspace{0.4cm}

{\em $^a$D\'epartement de Physique Th\'eorique and Centre for
  Astroparticle Physics (CAP), \\ Universit\'e de Gen\`eve, 24 quai E. Ansermet, CH-1211 Geneva, Switzerland}

{\em $^b$CERN, Theoretical Physics Department, Geneva, Switzerland}

{\em $^c$Perimeter Institute for Theoretical Physics, \\ 31 Caroline St.~N., Waterloo, Ontario N2L 2Y5, Canada
}

\end{center}

\vskip 1cm

\setlength{\baselineskip}{0.2in}

\centerline{\bf Abstract}
\vskip .1in

\noindent
The detection of primordial gravitational waves would not only have extraordinary implications for our understanding of early cosmology, but would also give non-trivial constraints on Standard Model parameters, under the assumption that no new physics enters below the Higgs instability scale. We study the resulting bounds on the top quark mass and the strong coupling constant, discussing their theoretical uncertainties and their robustness against changes in other parameters. 

\noindent

%\tableofcontents

%\end{abstract}
\end{titlepage}

\vskip .4in
\noindent

\section{Introduction}

The combination of the Planck 2018 and the BICEP2/Keck Array BK14 data give a tight upper bound on the tensor-to-scalar ratio $r$ (at the pivot scale $k=0.002$ Mpc$^{-1}$)~\cite{planck18}
\be
\label{eq: r Planck}
r< 0.064\,\,\,\,\,\,(95\% \,\,{\rm CL}) \, .
\ee
Recently, the BICEP2 and Keck Array Collaborations have published an equivalent
upper bound on $r$ at the pivot scale of $k=0.05$ Mpc$^{-1}$ \cite{south}
\be
\label{eq: r bickeck}
r< 0.06\,\,\,\,\,\,(95\% \,\,{\rm CL}) \, .
\ee
The amplitude of the tensor mode power spectrum is of paramount importance in cosmology since it is directly proportional to the vacuum energy density driving inflation (see {\it e.g.} ref.~\cite{lrreview}). In terms of  the  Hubble rate $H$ during inflation~\cite{planck18} 
\be
\frac{H}{M_{\rm Pl}}
=1.06\times 10^{-4}\, r^{1/2},
\label{Hrrel}
\ee
(where $M_{\rm Pl}$ is the reduced Planck mass) the  limits in eqs.~(\ref{eq: r Planck})--(\ref{eq: r bickeck})  translate into the upper bound 
\be
H< 6\times 10^{13}\,\,{\rm GeV}\, .
\label{Htyp}
\ee
Future observational efforts will be focused on detecting the primordial tensor modes through the B-mode polarisation  of the CMB anisotropies and both the 
US S3/S4 ground-based program \cite{stage41} and the LiteBIRD satellite \cite{litebird} are expected to deliver results at the end of the next decade.  
Forecasts for the measurement of the tensor-to-scalar ratio $r$  indicate that detection of the primordial B-mode polarisation is possible only if  
\be
\label{eq: r min}
r \simgt 5\times 10^{-4}\, ,
\ee
which corresponds to
\be
H \simgt 6 \times 10^{12}\,\,{\rm GeV}\, .
\ee
The value in \eqref{eq: r min} is admittedly optimistic and takes  into account noise sensitivity only. Once polarised dust is taken into consideration, the estimate on the reach in $r$ weakens by a factor of two~\cite{fut}. 
It has been suggested~\cite{book} that the intensity pattern of 21-cm radiation from the dark ages could provide a much more sensitive probe of the tensor-to-scalar ratio, in principle down to values as small as $10^{-9}$. However, due to the low frequency of the signal, such a formidable sensitivity requires
a very futuristic experiment.  An  interferometer with  a few hundred kilometer baseline  would  constrain $r$ to the level of $10^{-3}$ \cite{moon},  while $r\sim 10^{-6}$ could be reachable with an array extending over  a large fraction of the Moon's surface.

Limiting ourselves to optimistic (but realistic) prospects for the near future, we will consider \eqref{eq: r min} as the range of experimental sensitivity. This leads to a  window of opportunity of about one order of magnitude for determining the Hubble rate during inflation $H$ through detection of primordial gravitational waves: $6 \times 10^{12}\,\,{\rm GeV} \simlt H \simlt 6 \times 10^{13}\,\,{\rm GeV}$. If a positive detection is made in the future, the result can be used to derive useful constraints on Standard Model (SM) parameters, under the assumption that no new physics modifies the Higgs potential up to high scales. The derivation of these constraints and the discussion of their theoretical uncertainties is the subject of this paper.

These constraints originate from the observation that the SM Higgs potential develops an instability  at large field values which, for the current central values of the Higgs and top masses and the strong coupling constant, occurs around $10^{12}$ GeV \cite{instab2,instab,buttazzo,GD}.  At that scale, the Higgs quartic coupling $\lambda$ changes sign, suddenly driving the Higgs potential to become negative. 
As a consequence, the SM electroweak vacuum does not correspond to minimum energy, but to a metastable state. 

While our current SM electroweak vacuum is safe against both quantum  tunnelling in flat spacetime \cite{instab,buttazzo,Andreassen:2017rzq} and thermal fluctuations in the early universe \cite{espinosa}, the same may not be concluded for a period of primordial inflation. 
Even accepting the optimistic assumption that the inflationary stage starts with the SM Higgs sitting close to the origin, unavoidable quantum excitations of the Higgs itself will move its classical value away from its initial position. 
By performing a random walk caused by kicks of size $\pm H/2\pi$ at each Hubble time, the classical value of the Higgs can reach the top of the SM potential and fall deep into the unstable region~\cite{espinosa,cosmo2,Arttu}. 
If so, when inflation ends, the universe will be populated by anti-de Sitter (AdS) patches which  are lethal for our universe as they  grow at the speed of light engulfing our observable universe \cite{tetradis}. 

The value of the Hubble rate plays a critical role in the dynamics of the SM Higgs during inflation, as it determines the size of field fluctuations. This provides the exciting connection between the existence of tensor modes and the fate of the electroweak vacuum, which is at the core of the constraints on SM parameters derived in this paper. This connection allows us to ask questions such as: will a future detection of tensor modes tell us that our current electroweak vacuum is unstable under inflation? What is the range of SM parameters that can avoid the AdS catastrophe? Is the stability of the SM incompatible with a future detection of tensor modes?

By addressing these questions, we will show that the measurement of $r$ leads to a combined bound on the top mass and the strong coupling constant. This bound has only a weak dependence on the Higgs mass, but is critically sensitive to the Higgs-curvature coupling and the reheating temperature after inflation. Because of the relatively narrow range of $H$ that can be explored experimentally ($6 \times 10^{12}~{\rm GeV}\simlt H \simlt 6 \times 10^{13}~{\rm GeV}$) and the slow (logarithmic) evolution of $\lambda$ with the field value, the actual measured value of $r$ does not influence significantly the numerical bound on the SM parameters. Positive evidence for $r$ is all that matters for establishing quantitatively the bounds. 

The paper is organised as follows. In section 2 we discuss the Higgs potential during inflation, while in section 3 we describe the procedure to calculate the Higgs survival probability 
taking into account the post-inflationary evolution. Section 4 contains our results, and our conclusions are given in section 5.

%%%%%%%%%%%%

\section{The Higgs potential during inflation}

The first ingredient needed for our analysis is the Higgs potential during inflation. For the contribution in flat space, we employ the SM Higgs effective potential given in ref.~\cite{buttazzo}, containing up to two-loop corrections in Landau gauge with the SM parameters evolved to large energies with full three-loop RGE precision. We focus on observables that are gauge independent~\cite{schwartz}, giving us the freedom to choose the gauge-fixing parameters of the effective action \cite{tetradis,GD}. 
Moreover, the renormalisation scale will be taken as 
\be
\label{n}
\mu\approx \sqrt{h^2+12H^2},
\ee
($h$ being the value of the Higgs field)  which represents the best choice when dealing with the ultraviolet contribution of the quantum degrees of freedom  coupled to the SM Higgs in  de Sitter space~\cite{scale}. 

\begin{figure*}[t!] \centering
\includegraphics[width=.45\textwidth]{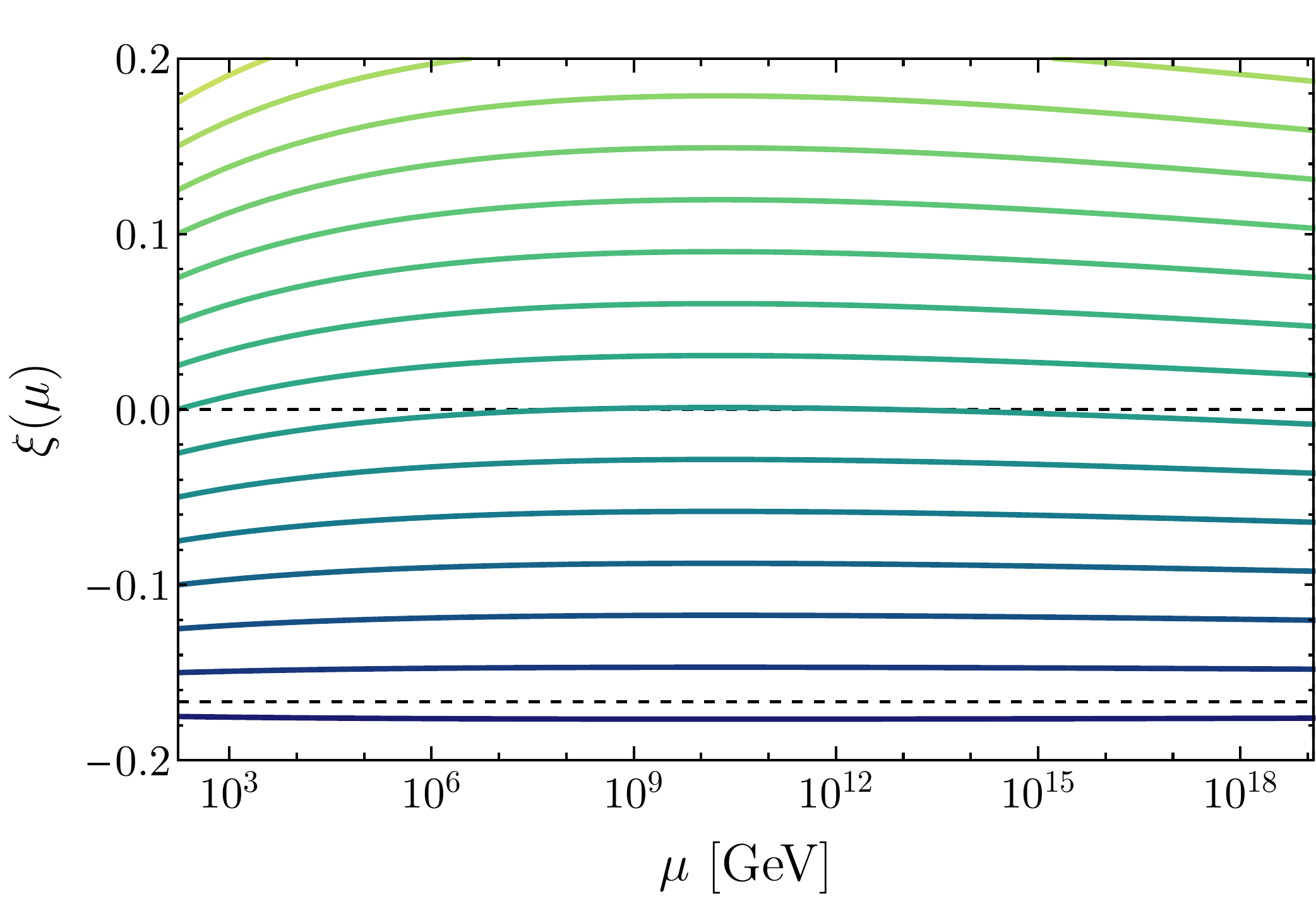}
\includegraphics[width=.4725\textwidth]{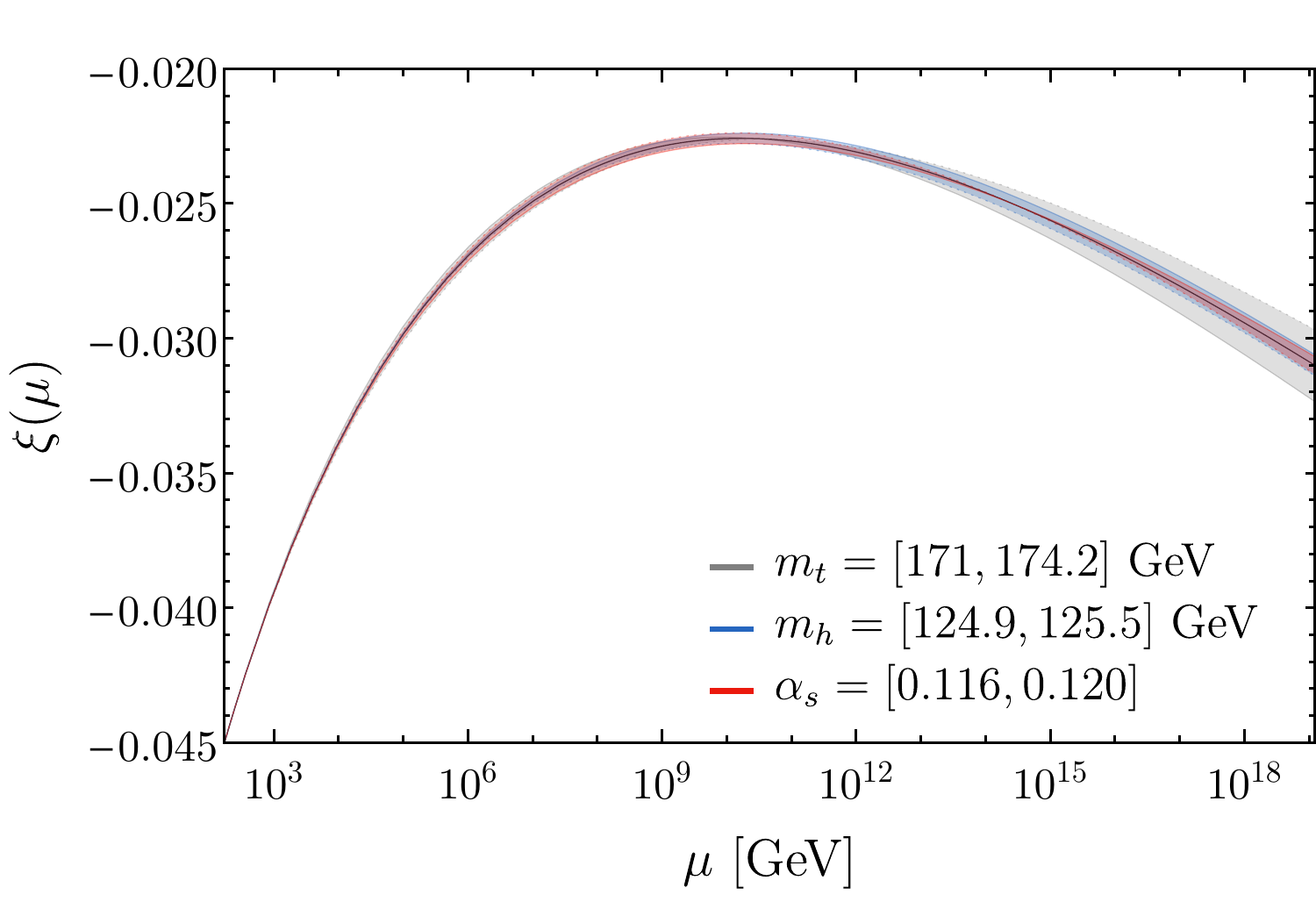}
\caption{\textit{Left:} Running of $\xi(\mu)$ with different initial conditions at $\mu = m_t$ and for the central values of the SM parameters. \textit{Right:} Running of $\xi(\mu)$ for $\xi(m_t)=-0.045$ varying the SM parameters as indicated in the figure.
 }
\label{fig:xi}
\end{figure*}

The leading terms of the Higgs potential during inflation are given by~\cite{tetradis,scale}
\be
\label{pot1}
V_{\rm SM}(h,\mu)=e^{4\Gamma(\mu)}\frac{\lambda(\mu)}{4} h^4-6H^2\xi(\mu)e^{2\Gamma(\mu)}h^2+\alpha(\mu)H^4 \, .
%+V_{\rm top}(h,\mu).
\ee
Here $\lambda(\mu)$ is the running Higgs quartic coupling at NNLO precision (in the accuracy of our analysis), and  
 $\Gamma$ is the (integral of the) gauge-dependent anomalous dimension of the Higgs field, related to the wave-function $e^{2\Gamma}$, which multiplies the kinetic term of the Higgs in the effective action expanded in derivatives~\cite{tetradis}. Moreover, 
$\xi$ is the 
Higgs-curvature coupling (defined such that $\xi=-1/6$ corresponds to a conformally-invariant coupling), whose running is shown in fig. \ref{fig:xi}. 
The third term in eq.~\eqref{pot1} describes the effect of higher-curvature interactions in de Sitter space with a coupling constant $\alpha(\mu )$, whose RG flow is given by~\cite{scale}
\be
\frac{{\rm d}\alpha}{{\rm d}\ln\mu}=\frac{1}{16\pi^2}\left(288\, \xi^2+96\, \xi-\frac{1751}{30}\right) \, .
\label{alpeq}
\ee
Note from eq.~\eqref{alpeq} that $\alpha$ is renormalised additively and therefore its boundary condition at a given scale $\mu_0$ contributes to the potential only with a field-independent constant $\alpha (\mu_0 )H^4$. Thus, the physically relevant contribution from $\alpha$ is fully described by the value of $\xi$ and does not introduce any additional free parameter.

The potential (\ref{pot1}) is augmented with a contribution  from the real top quarks that are unavoidably created during inflation, generating a non-vanishing condensate $\langle \overline t t\rangle$ \cite{f}. 
This is the quantity which appears in the equations of motion for the Higgs field and is relevant for us, therefore we compute it directly and insert  it in the equations of motion. 
A more detailed procedure to compute the full fermion effective potential in de Sitter space can be found in \cite{fermionEffPotdS}. In practice, this contribution to the Higgs potential turns out to be always largely subodminant, so that we do not need a very refined estimate.
%Essentially, the top quarks produced during inflation generate a non-vanishing condensate $\langle \overline t t\rangle$ which is time-independent since  de Sitter is a maximally symmetric spacetime with no preferred coordinates (we include the color factor neglected in ref.~\cite{f})
The top condensate $\langle \overline t t\rangle$, which is time-independent since  de Sitter is a maximally symmetric spacetime with no preferred coordinates, can be written as (we include the color factor neglected in ref.~\cite{f})
\be
\langle \overline t t\rangle\simeq 12 \int\frac{\d^3 q}{(2\pi)^3}\frac{m_t(h)}{\sqrt{q^2+m^2_t(h)}}|\beta_q|^2 \, .
\ee
Here $q(t)=k/a(t)$ is the physical momentum ($k$ being the comoving momentum), $m_t(h)=y_t\,  h/\sqrt{2}$ is the top mass in the Higgs background ($y_t$ being the top Yukawa coupling) and $\beta_q$ is the Bogoliubov coefficient which can be approximated by \cite{pr}
\be
\label{beta}
\beta_q(t)\approx \frac{1}{2}\int_{-\infty}^t\d t'\frac{m_t(h)\, q\,  H}{q^2+m^2_t(h)}e^{-2 i \int_{-\infty}^{t'}\d t''\sqrt{q^2+m^2_t(h)}} \, .
\ee
If the phase of the exponential in eq.~(\ref{beta}) is  of order unity or larger, then the oscillatory behaviour will damp $\beta_q$. 
Its  final magnitude depends therefore on the size of $q$ with respect to the rate of change of the rest of the integrand, which is given by $|\dot q/q|=H$ (the rate of change of the Hubble rate and the Higgs field are negligible). 
Consider first the production of relativistic particles, with $q\gsim m_t$. In this case the frequency of the oscillations is approximately equal to the physical momentum and particle production is (exponentially) suppressed for $q\gsim H$, while remains unsuppressed for $q\lsim H$ (which requires $m_t\lsim H$). 
Therefore, in the relativistic case, we find
\be
\label{beta2}
\beta_q(t)\Big|_{ {\rm rel}}\approx \frac{1}{2}\int_{k/H}^{k/m_t}\frac{\d a'}{a'H}\frac{m_t  H}{k}a'\simeq\frac{1}{2}\left(1-\frac{m_t}{H}\right).
\ee
In the non-relativistic case, $q\lsim m_t$, the frequency of the oscillations is approximately equal to the mass $m_t$
and,  again, the non-oscillatory part of the integrand is $H$. 
Particle production is unsuppressed for $m_t\lsim H$ and we find
\be
\label{beta3}
\beta_q(t)\Big|_{ {\rm non-rel}}\approx \frac{1}{2}\int_{k/m_t}^\infty \frac{\d a'}{a'H}\frac{k H}{a' m_t}\simeq\frac{1}{2}.
\ee
All in all, in the regime $m_t\lsim H$, the largest contribution to the top condensate comes from the relativistic infra-red modes during inflation with  $q\lsim H$
\be
\langle \overline t t\rangle\simeq 12 \int\frac{\d^3 q}{(2\pi)^3}\frac{m_t(h)}{4q}\simeq \frac{3m_t}{2\pi^2}\int_{m_t}^H\d q\, q\simeq \frac{3m_tH^2}{4\pi^2},
\ee
while the contribution from the non-relativistic modes gives a top condensate which scales like $m_t^3$. 

Including the exponential suppression that occurs in the regime $m_t > H$ and extrapolating the overall coefficient from the numerical results of ref.~\cite{long}, we end up with the following analytical estimate for the contribution to the effective potential due to top quanta  
\be
V_{\rm top}(h,\mu)= m_t \langle \overline t t\rangle \simeq  3\times 10^{-3}\, H^2\, m^2_t(h) \, e^{2\left[ \Gamma(\mu)-m_t(h)/H\right]}
\ee
for  $m_t(h)\gsim 0.1\, H$. 
This is analogous to the   corrections to the SM Higgs potential in a thermal plasma. 
Contrary to what found in ref.~\cite{f},  the condensate is  exponentially suppressed at high masses as the arguments above show. This is confirmed both analytically, by applying the stationary phase method \cite{chung}, or numerically \cite{long}.

Analogous considerations can be applied to the contribution from gauge bosons. Transverse degrees of freedom behave like conformally coupled scalar fields and their production is, like for fermions, proportional to their mass. We find
\begin{equation}
V_{\rm gt}(h,\mu) \ \simeq \  10^{-3}\,  H^2 e^{2\Gamma(\mu)} \left[4 m_W^2(h)  e^{-2m_W(h)/H}
  +  2m_Z^2(h)  e^{-2 m_Z(h)/H}\right],
\end{equation}
where $m_W^2(h)=g^2 h^2/4$, $m_Z^2(h)=(g^2+g'^2)h^2/4$, with $g$ and $g'$ being the gauge couplings of $SU_L(2)$ and $U_Y(1)$, respectively. 

Longitudinal  degrees of freedom behave like minimally coupled scalar fields \cite{long} and 
their contribution to the potential comes from their variance, proportional to (taking, as an example, the case of $W$-bosons)
\be
\left(\frac{H}{2\pi}\right)^2\int{\rm d}\ln k\,\left(\frac{k}{H}\right)^{3-2\nu_W}, \,\,\,\,\nu_W=\sqrt{9/4-m_W^2/H^2}.
\ee 
 This gives rise to a contribution to the Higgs potential of the form 
$m_W^2H^2/4\pi^2(3-2\nu_W)$ for $m_W(h)<3H/2$ and exponentially suppressed otherwise. 
Since the survival condition imposes  $H$ to be smaller than $h_{\rm max}$, we expect  the contribution of the longitudinal gauge bosons to be negligible in the field region  of interest. 
Notice, moreover, that the contribution from gravitationally produced tops and gauge bosons is typically subdominant to the contribution from non-minimal Higgs coupling to gravity, which cannot be set to zero at all scales, since $\xi =0$ is not stable under RG evolution.

\section{The Higgs survival probability}

\subsection{Inflationary dynamics}

The next ingredient of our analysis is the calculation of the Higgs survival probability, defined as the probability that the Higgs does not end up in the AdS region at the end of inflation. Taking a constant Hubble rate $H$ during inflation and trading time with the number of $e$-folds ($\d N=H\d t$), we can write the gauge-independent  Langevin equation of motion~\cite{starob} of the long wavelength modes of the 
Higgs field \cite{tetradis,GD}
\be
\label{Langevin}
e^{\Gamma}H\frac{\d h}{\d N} + \frac{e^{-\Gamma}}{3H}\frac{\d V(h)}{\d h}=\eta(N)\, .
\ee
This equation is valid as long as the effective Higgs mass is smaller than $3H/2$. 
Indeed, the solution to the Klein-Gordon equation in de Sitter depends on the mass only through the quantity $(9/4-m^2/H^2)$, so that the effect of the mass is negligible if it is smaller than $3H/2$.
The possible effect of a small mass in the power spectrum of the perturbations and therefore in the noise of the Langevin equation is to provide an IR cut-off to the perturbations, an effect which is well captured in the distribution probability for the Higgs (see for instance \cite{tetradis}).
In particular, for the Higgs mass coming from the non-minimal gravitational coupling, see eq.~\eqref{pot1}, this is negligible with respect to $3H/2$ for $\xi> -3/16$.  Smaller values of $\xi$ lead to a complete damping of Higgs fluctuations and the field remains frozen near the origin. 
The Higgs mass induced by its self-coupling is equal to $3\lambda h^2$, and is negligible for $h<H\sqrt{3/(4\lambda)}$, a condition which is well satisfied in the range of Higgs excursions and Hubble rates we consider.
Accounting for the wave-function renormalisation ({\it i.e.} the $e^\Gamma$ factors) in eq.~\eqref{Langevin} allows for a significant reduction in the gauge dependence of 
$h_{\rm max}$\footnote{Here and in the following  we define  $h_{\rm max}$ as the value of the Higgs at the maximum of the   full  potential during inflation, see discussion below. We warn the reader that this definition differs from ref.~\cite{tetradis}, where $h_{\rm max}$ indicates the value of the Higgs at the maximum of the present-day potential \cite{tetradis,GD}. We will call such a value $\Lambda$.}.
Finally, $\eta$ is a Gaussian random noise with
\be
\langle\eta(N)\eta(N')\rangle=\left(\frac{H^2}{2\pi}\right)^2\delta_D(N-N').
\ee
Once the Langevin equation \eqref{Langevin} is solved for many stochastic realisations, we can construct the probability $P(h,N)$ that the Higgs is at a given value $h$  at a given number of $e$-folds $N$. From this, we can compute the survival probability $P_{\rm surv}$ ({\it i.e.} the probability that the Higgs does not surmount the barrier)
\be
P_{\rm surv}(N)=\int_{-h_{\rm end}}^{h_{\rm end}}\,\d h\, P(h,N) \, .
\label{eq: survival prob}
\ee
Notice that this quantity is gauge-independent \cite{tetradis,GD}. Here $h_{\rm end}$ is defined to be the largest value of the Higgs field at the end of inflation for which, during the  subsequent post-inflationary evolution,   the Higgs field has time
enough to roll down to the safe region  $h<\Lambda$, being $\Lambda$ the location of the maximum of the current Higgs potential $\lambda(h)h^4/4$.
Since there are  $\approx e^{3N}$ ($N\sim 60$ being the number of $e$-folds till the end of inflation corresponding to our current horizon) causally independent regions that are generated  during inflation and end up in our  observable universe today, we will impose the condition that the probability for the Higgs to remain trapped in the AdS region is less than $ e^{-3N}$ or, equivalently,
\be
1-P_{\rm surv}(N=60) < e^{-180}\approx 10^{-78}\, .
\label{eq: surv N=60}
\ee
Here we have (optimistically) assumed that the Higgs is sitting at the current electroweak vacuum about $60$ $e$-folds before the end of inflation\footnote{The fact that after $N$ $e$-folds
one obtains the relation $\langle h(\vec x)h(\vec y)\rangle\simeq \langle h^2(\vec x)\rangle$ over a volume of the order of $H^{-3} e^{3N}$ \cite{vilenkin} might give the impression that the condition (\ref{eq: surv N=60}) is too restrictive and that the Higgs has a common value over patches much larger than a single Hubble volume. However, such relation only says that  over a volume  of order of $H^{-3} e^{3N}$  one can find any value of the Higgs field
between 0 and the square root of the variance in any Hubble volume $H^{-3}$. The condition (\ref{eq: surv N=60}) is therefore the correct criterion to assure that, in regions of
Hubble volume $H^{-3}$, the Higgs is not in AdS.}. Given that the Higgs undergoes a stochastic motion with steps $H/2\pi$, its precise initial condition is not important, as long as the Higgs starts below the instability scale.

\subsection{Post-inflationary dynamics}

The survival probability depends on the Higgs value $h_{\rm end}$. We suppose that, after the end of inflation, the universe enters a matter-dominated phase during which the inflaton  field responsible for inflation oscillates around the minimum of its potential. The Hubble rate therefore scales like $H_{\rm m}=H/a^{3/2}$, where $a$ is the scale factor and we have assumed the end of inflation to be at $a=1$. Meanwhile, the inflaton field decays
and the universe becomes more and more filled with relativistic degrees of freedom. The corresponding temperature has the following behaviour \cite{Giudice:2000ex}
\be
T(a)\simeq 1.3~ T_{\rm max}\,a^{-3/8}(1-a^{-5/2})^{1/4},
\ee
where 
\be
T_{\rm max}\simeq 0.54\left(\frac{HM_{\rm Pl}T_{\rm RH}^2}{g_*^{1/2}}\right)^{1/4}\, .
\ee
 $T_{\rm RH}$ is the reheating temperature and $g_*$
 the effective number of relativistic degrees of freedom.  Within a Hubble time, the temperature raises up from zero to  a maximum value $T_{\rm max}$ 
and then decreases as $a^{-3/8}$. During this stage, entropy from the decay of the inflaton field is released. When the age of the universe becomes comparable to the lifetime of the inflaton, the temperature reaches the value $T_{\rm RH}$. At this stage the universe becomes radiation-dominated and the temperature starts scaling as $a^{-1}$.

During this phase, the Higgs potential acquires additional corrections due to thermal effects:
\be
V_T(h)\simeq \kappa \, T^2\,\frac{h^2}{2}e^{-h/(2\pi T)},\,\,\,\,\kappa\simeq 0.12.
\ee
This approximate expression is obtained within the high temperature limit (hence the exponential factor), and the prefactor is computed with a numerical fit which is accurate for $h \lesssim 10 T$ and includes the effect of ring resummation \cite{Espinosa:2017sgp}.

Our strategy to determine $h_{\rm end}$ is the following. At the end of inflation, during the matter-dominated phase, but before reheating is completed, the equation of motion
of the Higgs is
\be
\frac{\d^2 h}{\d a^2}+\frac{5}{2a}\frac{\d h}{\d a}+\frac{a}{H^2}V'=0.
\ee
The solution of this equation is $h(a)=h_{\rm end}f(a)$, such that $f(a=1)=1$. Keeping only the mass terms and neglecting the quartic coupling we take 
\be
V\simeq V_2(h,a)=\left(-\frac{3}{ a^3}\xi H^2+\kappa \frac{T_{\rm max}^2}{ a^{3/4}}\right)\frac{h^2}{2}\, .
\ee
Both the  potential and the Higgs are   decreasing with time and one has to insure that at the scale factor $a_{\rm max}$ at which
\be
V_2(\Lambda,a_{\rm max})=\frac{\lambda(\Lambda)}{4}\Lambda^4,
\ee
the Higgs has already passed the value $h=\Lambda$. This provides the relation
\be
h_{\rm end}\simlt  \Lambda f^{-1}(a_{\rm max}).
\ee
It is intuitive to understand that, if $T_{\rm max}$ is large enough, values of $h_{\rm end}$ larger than $h_{\rm max}$ during inflation may be rescued as long as they are
smaller than the value of the Higgs field $h_{\rm cl}\simeq (-3/2\pi\lambda)^{1/3}H$ above  which the classical dynamics dominates over the quantum one. Indeed, if this is not the case, the Higgs would have rolled down towards the instability even during inflation. 

\section{Results}

The calculation of $P_{\rm surv}$ depends on the Hubble rate during inflation $H$, the reheating temperature $T_{\rm RH}$, the Higgs-curvature coupling $\xi$, and the three most relevant SM parameters: $m_t$ (the top quark mass), $m_h$ (the Higgs mass),  $\alpha_s$ (the strong coupling constant). Thus, the survival condition in eq.~\eqref{eq: surv N=60} can be read as a bound on these parameters.

The value of $H$ can be traded with the tensor-to-scalar ratio $r$ through eq.~\eqref{Hrrel}. The Higgs-curvature coupling $\xi$ will be treated as a free parameter and, in all our plots and numerical considerations, it is evaluated at the renormalisation scale $6\times 10^{13}$~GeV, corresponding to the typical Hubble rate during inflation in case of a detection of $r$, see eq.~\eqref{Htyp}. For the Higgs mass and the strong coupling constant we use $m_h=(125.18\pm 0.16)$~GeV and $\alpha_s (m_Z)=0.1181\pm 0.0011$ ~\cite{pdg}. 

One of the major sources of uncertainties in our analysis is associated with $m_t$. The top quark mass can be extracted from hadron collider measurements by fitting $m_t$-dependent kinematic distributions to Monte Carlo simulations, using full or partial reconstruction of the ${\bar t}t$ decay products. These are usually called ``direct measurements" and
the PDG~\cite{pdg} quotes a world average of $m_t =173.0 \pm 0.4$~GeV, while it gives $m_t =173.1 \pm 0.9$~GeV for the ``pole mass" extracted from cross-section measurements. Recent combinations of different LHC channels~\cite{LHCt} give $m_t =172.8 \pm 0.7$~GeV (ATLAS) and $m_t =172.4 \pm 0.5$~GeV (CMS). Aside from the experimental error of these results, the main issue for us is how to relate the value of the top mass measured at the LHC with the parameter $m_t$ that enters our calculation of the Higgs potential, which is defined as the pole top mass, related in perturbation theory to the $\overline{\rm MS}$ running mass. (For a recent discussion of these issues, see ref.~\cite{nason}.) For a comparison with data, we will take an average of the recent ATLAS and CMS combinations and inflate the error to take into account the theoretical uncertainty on the value of $m_t$ that enters our calculation. Thus, we use $m_t =172.6 \pm 0.8$~GeV. We want to emphasise that this choice does not affect our results, but matters only for the visual comparison in the plots between the theoretical prediction and the experimental data. We also remark that this source of theoretical uncertainty could be bypassed in the future by using the direct measurement of the top Yukawa coupling $y_t$ from Higgs production in association with a top pair, since $y_t$ is the relevant parameter for the calculation of the Higgs potential and its extraction from data is free from any non-perturbative ambiguity. Present LHC measurements of $y_t$ are not sufficiently precise to be competitive with the direct measurements of $m_t$, but at a future 100 TeV collider a precision of 1\% or below is foreseeable~\cite{fcc}.  

To study the evolution of the Higgs field, for each set of the relevant six parameters ($m_t$, $m_h$, $\alpha_s$, $r$, $T_{\rm RH}$, $\xi$), we have solved $10^3$ independent realisations of the Langevin equation for $N=60$ $e$-folds to construct the survival probability and calculated the bound in eq.~\eqref{eq: surv N=60}.  We will present our result by identifying three regions in the plots. 
%\vskip 0.4cm
%\noindent
%
\begin{itemize} [label=\textcolor{verde}{\textbullet}]
\item  \textcolor{verde}{\it\bf Green.} 
This is the region allowed by the current bound on the tensor-to-scalar ratio $r$ in eq.~\eqref{eq: r Planck} and which will remain allowed independently of any future detection. 
%
%
%\vskip 0.4cm
%\noindent
%
\item[\textcolor{rosso}{\textbullet}] \textcolor{rosso}{\it\bf Red.}
This is the region that will be ruled out if future experiments detect $r$. 
%
%\vskip 0.4cm
%\noindent
%
\item[\textcolor{arancione}{\textbullet}]  \textcolor{arancione}{\it\bf Orange.}
This is the band that describes the bound on the underlying parameters, assuming detection of the tensor-to-scalar ratio somewhere in the range $5\times 10^{-4}<r<6.4 \times 10^{-2}$.  
\end{itemize}
It is convenient to separate the discussion into three parts, corresponding to different ranges of $\xi$ and different physical situations.

\subsection{Case {\boldmath $ \xi\lsim -3/16$} }
In this case the Higgs perturbations are damped since the  Higgs mass squared at the origin  during inflation is larger than $9H^2/4$. The Higgs field remains frozen at the origin and inflation does not trigger any Higgs instability.

\subsection{Case {\boldmath $ -3/16 \lsim \xi \lsim 0.1$} }

In this range the stochastic fluctuations of the Higgs are active and may destabilise the Higgs vacuum. The upper end of the range of $\xi$ ($\simeq 0.1$) comes from the following considerations.
For positive values of $\xi$, the origin of the Higgs potential is destabilised during inflation. Disregarding for the moment the running of the couplings, a new minimum is created at $\langle h \rangle \simeq H \sqrt{12\, \xi /\lambda}$, if the corresponding Higgs value is below the instability scale. (We will discuss in sect.~\ref{secxi01} the condition for this to happen.) At this minimum, the Higgs squared mass is about $24\, \xi H^2$. Therefore, for $\xi \simgt 3/32$, the Higgs field during inflation is frozen at the new minimum and quantum fluctuations are ineffective. In fact, we have checked numerically that the critical value of $\xi$, when accounting for the running of the couplings, is 
(10--30)\% larger than $3/32$ and therefore we have taken $\xi\simeq 0.1$ as a benchmark value. 
Here we discuss the case $\xi\lsim 0.1$, in which the Higgs field undergoes a stochastic motion during inflation, and we leave the opposite case for sect.~\ref{secxi01}.

\begin{figure}[h!] \centering
\includegraphics[width=.45\textwidth]{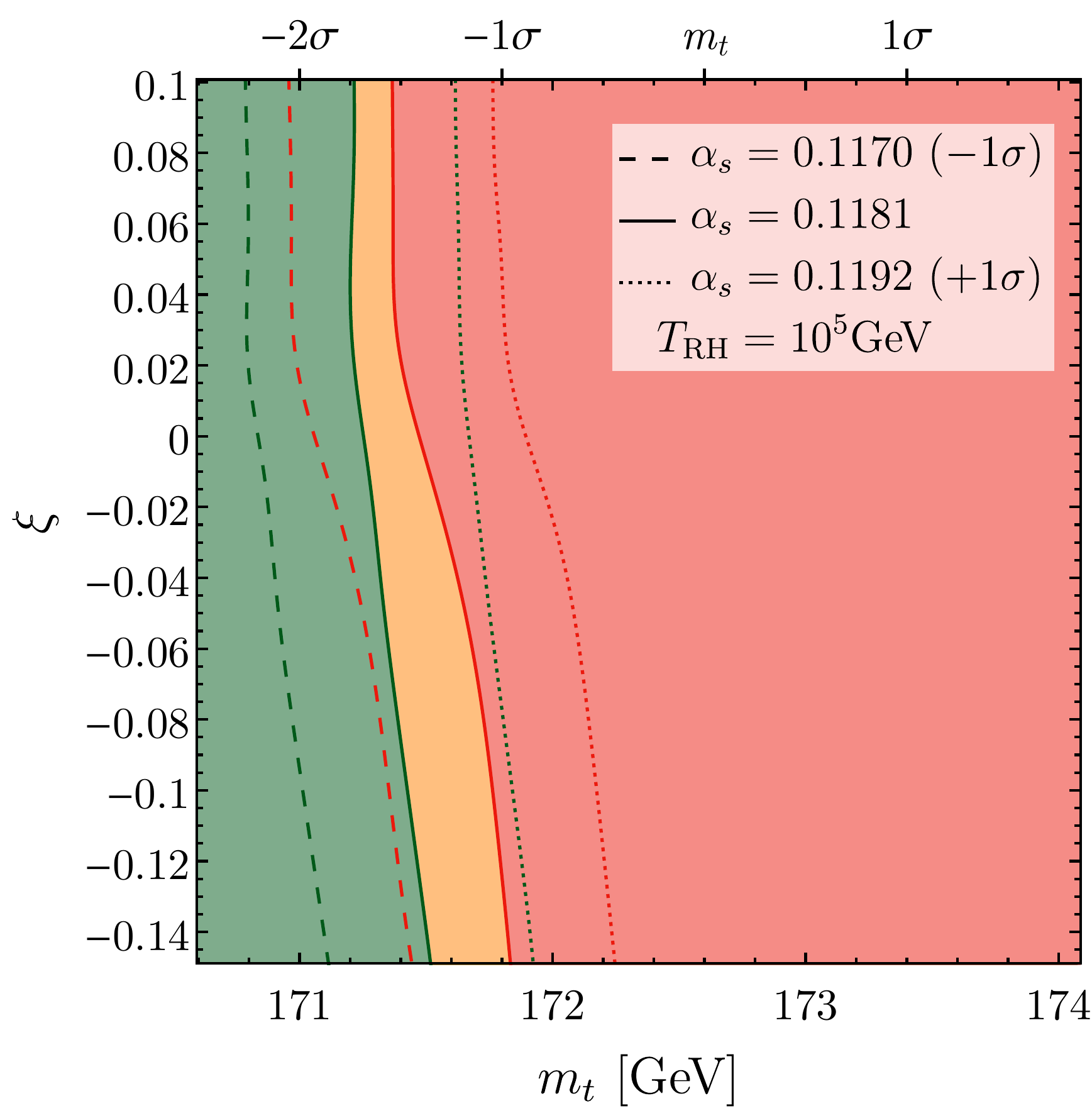}
\includegraphics[width=.45\textwidth]{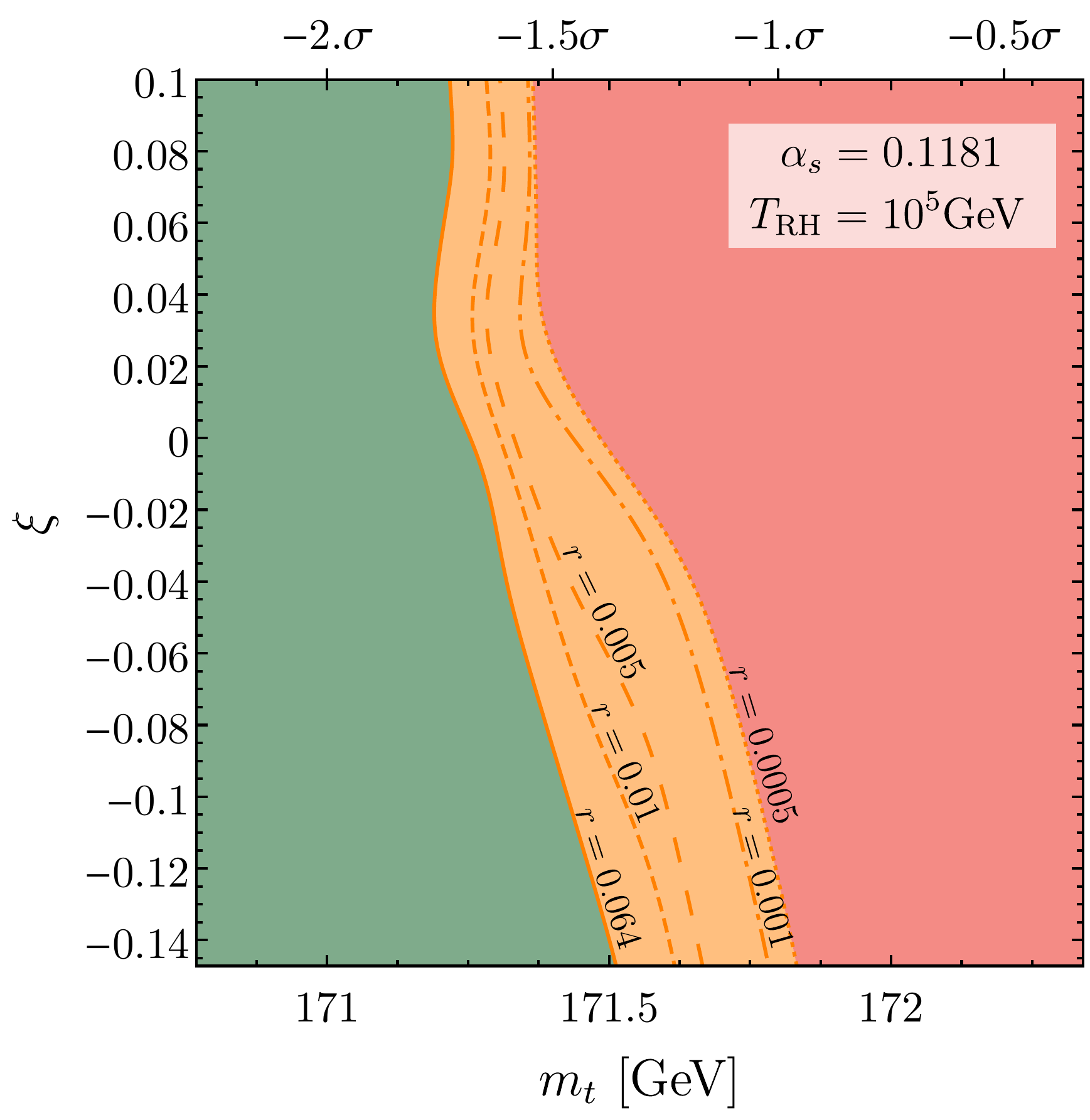}
\caption{
\textit{Left:} The three regions discussed in the text, in the $m_t$--$\xi$ plane, where $\xi$ is the Higgs-curvature coupling evaluated at the scale $6\times 10^{13}$~GeV. The figure illustrates the low $T_{\rm RH}$ regime, corresponding to the most stringent bounds. Dashed and dotted lines show how the region boundaries change with variations of $\alpha_s$. The Higgs mass is fixed at its central value. 
\textit{Right:}  Zoom over the orange band, fixing $\alpha_s$ to its central value. The different lines show the bounds corresponding to different measurements of the tensor-to-scalar ratio $r$.
 }
\label{fig: mt xi}
\end{figure}
\begin{figure}[h!] \centering
\includegraphics[width=.32\textwidth]{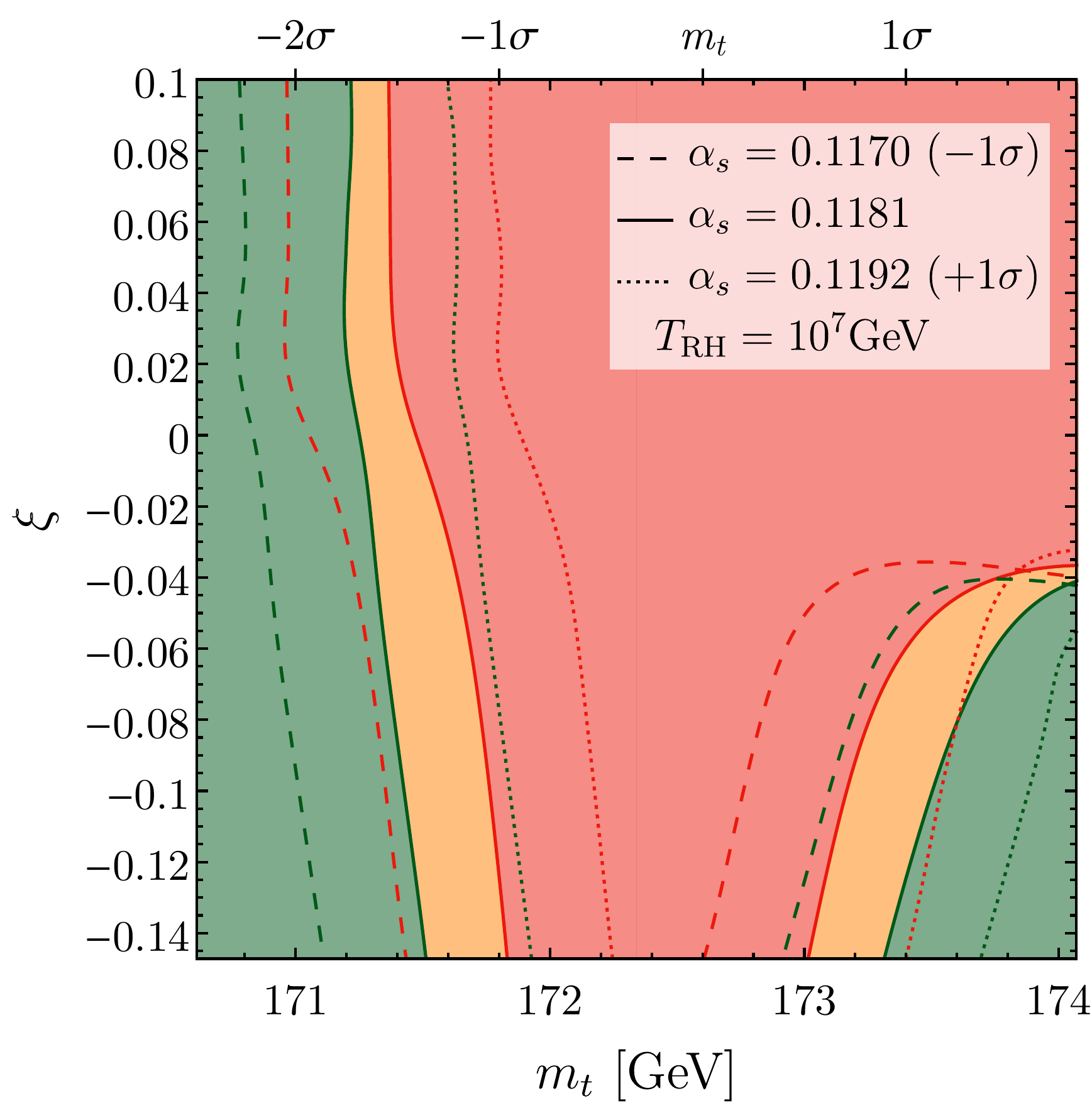}
\includegraphics[width=.32\textwidth]{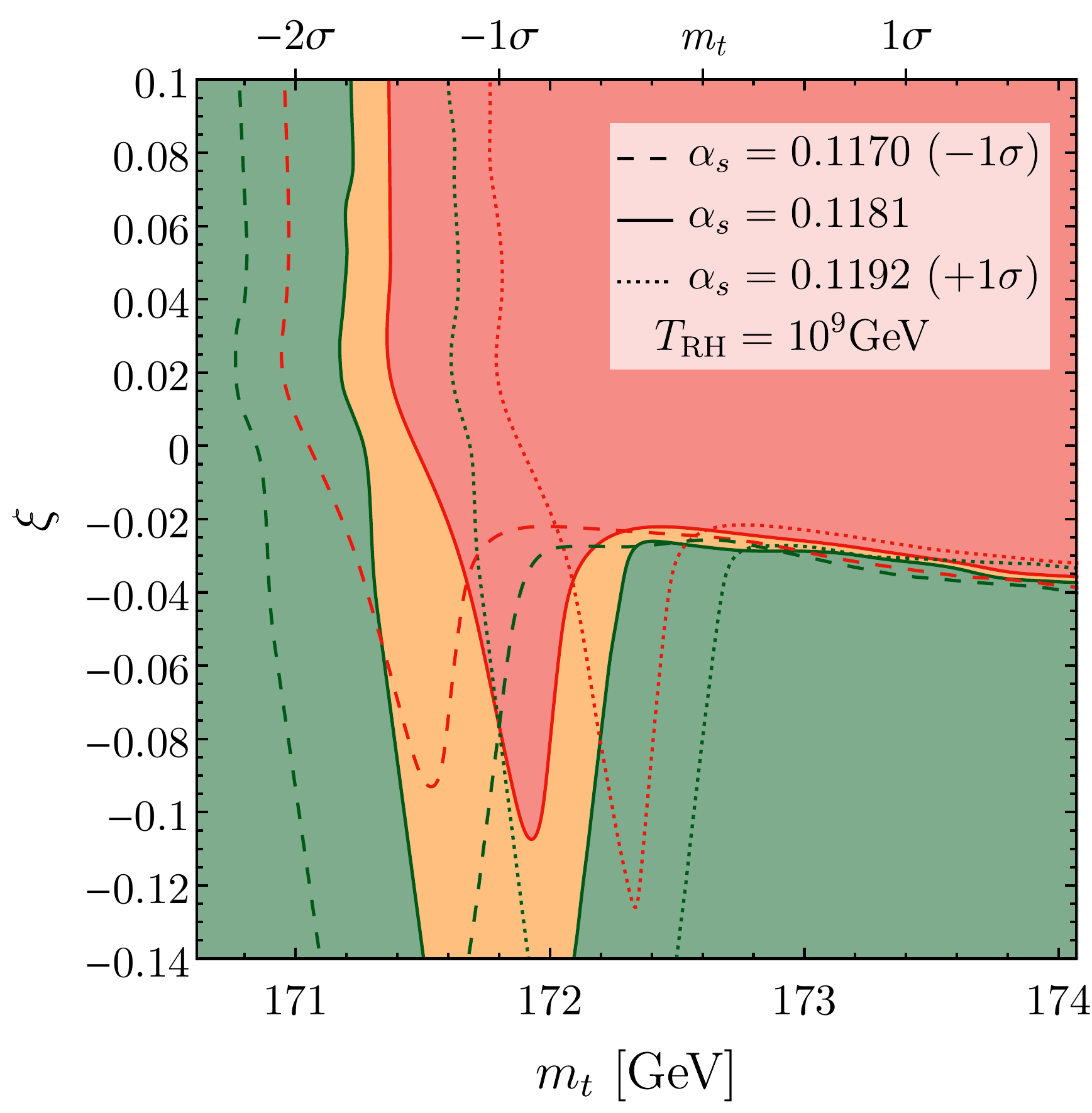}
\includegraphics[width=.32\textwidth]{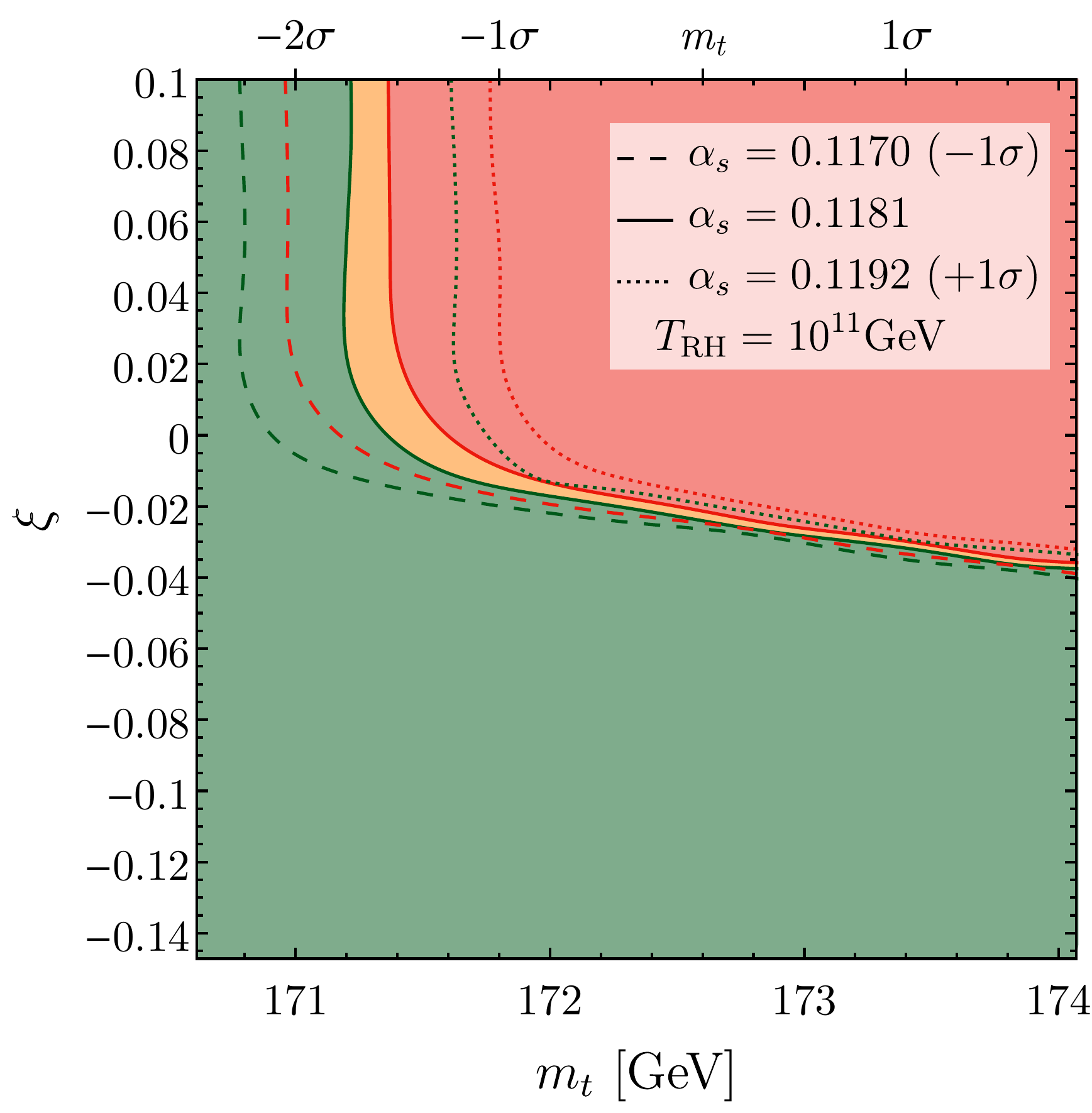}
\caption{
As in fig.~\ref{fig: mt xi}, for different values of $T_{\rm RH}$. 
 }
\label{fig: mt xi 2}
\end{figure}

In figs.~\ref{fig: mt xi} and \ref{fig: mt xi 2} we show the bound coming from a hypothetical measurement of $r$  in the plane $m_t$--$\xi$ for three choices of $\alpha_s$, the central value of $m_h$ and various  choices of $T_{\rm RH}$.
The larger the value of the measured $r$, the more stringent the bound, as shown in the zoomed region in the right plot in fig.~\ref{fig: mt xi}. 
For low values of $T_{\rm RH}$ (see fig.~\ref{fig: mt xi}), the bound on $m_t$ becomes very strong. Notice that the bound on $m_t$ is present for either signs of $\xi$. 
For  $\xi<0$, one could expect its positive contribution to the Higgs mass squared to sensibly alleviate the instability issue. However, in such a case the maximum of the Higgs potential is shifted towards values larger than $\Lambda$ and therefore the Higgs fluctuations may push the Higgs to values between $\Lambda$ and the new maximum during inflation. During preheating, the Higgs field might not successfully  roll down to the safe region of the potential after inflation.
%Obviously, for $\xi <-3/16$, when Higgs fluctuations during inflation are damped by the effective Higgs mass, the region of the parameter space turns green. 

%
\begin{figure}[h!] \centering
\includegraphics[width=.545\textwidth]{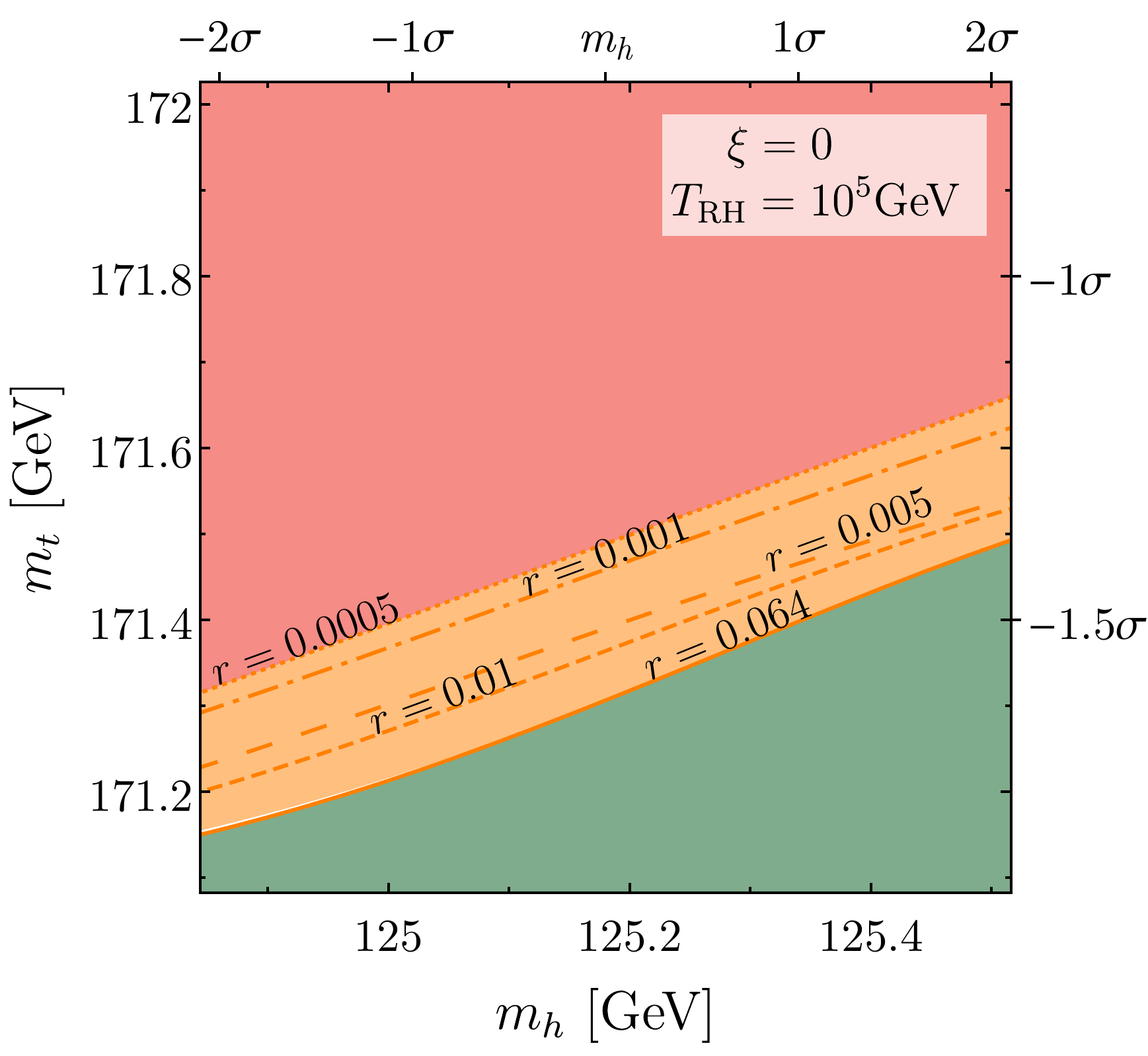}
\caption{As in fig.~\ref{fig: mt xi}, but showing the dependence on the Higgs mass $m_h$, while fixing $\alpha_s$ to its central value and the Higgs-curvature coupling (evaluated at the scale $6\times 10^{13}$~GeV) to $\xi=0$.}
\label{fig: mt mh}
\end{figure}
\begin{figure}[h!] \centering
\includegraphics[width=.545\textwidth]{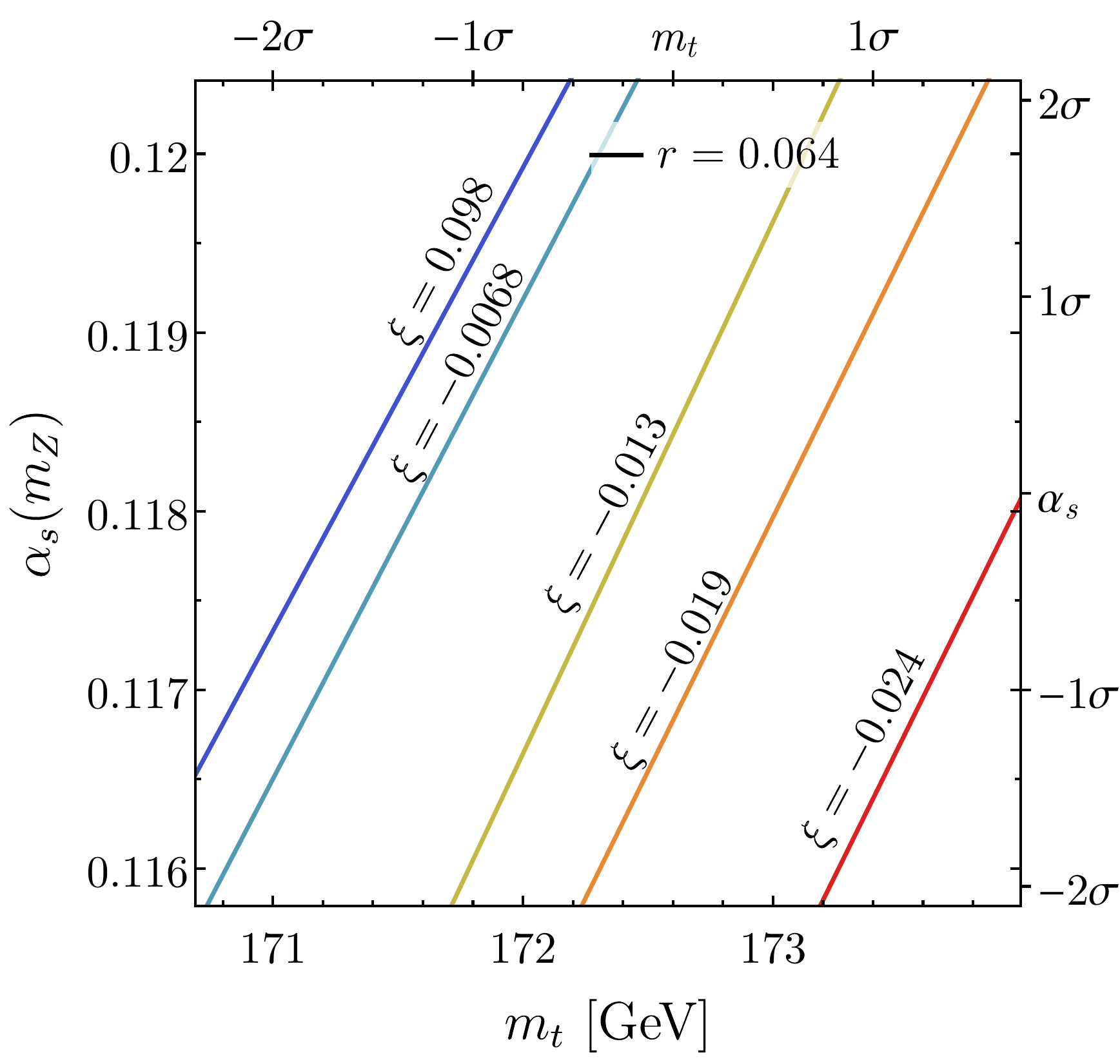}
\caption{Bounds on $m_t$, $\alpha_s$, $\xi$ from a positive detection of the tensor-to-scalar ratio $r$ (here $r$ is taken equal to $0.064$, but the bounds are largely insensitive to the value of $r$ within detection reach). The figure corresponds to the high $T_{\rm RH}$ regime ($T_{\rm RH}\ge 10^{13}$~GeV), which gives the most conservative bounds. For values of $\xi \simgt 0.02$ the lines overlap and become indistinguishable. The figure can be read in two ways. {\it (i)} For given $m_t$ and $\alpha_s$, the lines show the corresponding upper bound on $\xi$. {\it (ii)} For a given $\xi$, the corresponding line shows the bound on $m_t$ and $\alpha_s$, with the region to the right of the line excluded by the measurement of $r$.}
\label{fig: mt alpha xi}
\end{figure}
\begin{figure}[h!] \centering
\includegraphics[width=.45\textwidth]{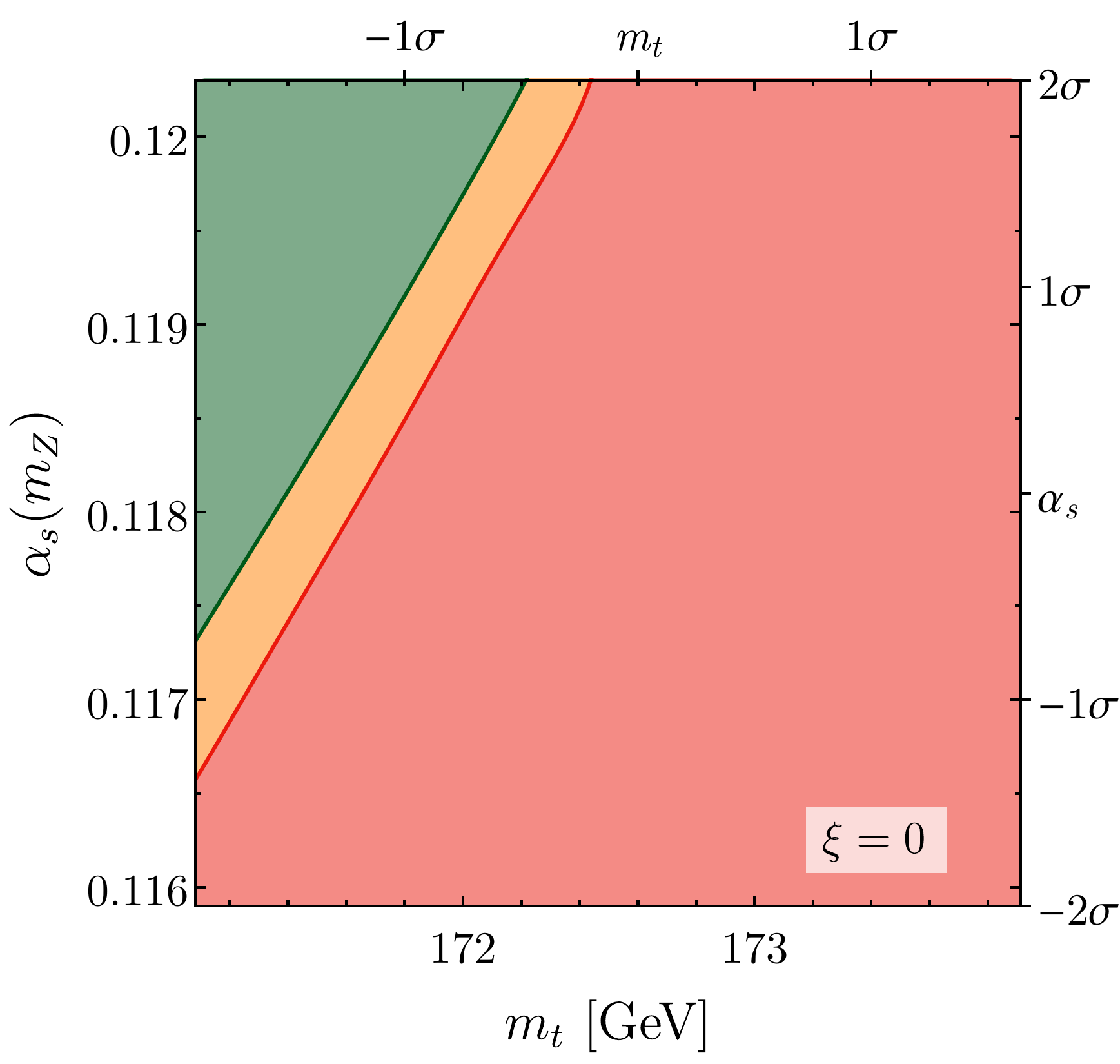}
\includegraphics[width=.45\textwidth]{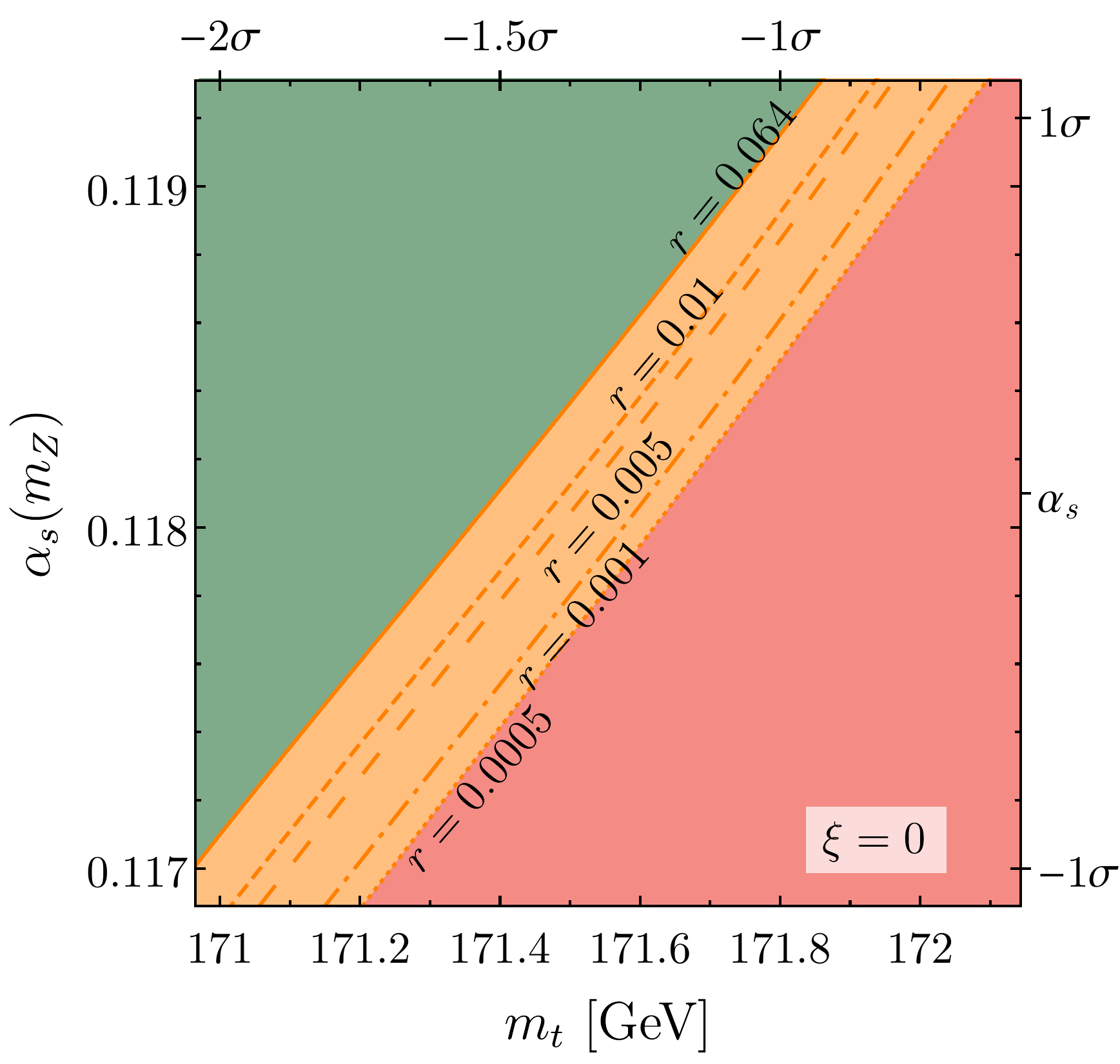}
\caption{As in fig.~\ref{fig: mt alpha xi}, for the special value $\xi =0$. The bound corresponds to the high $T_{\rm RH}$ regime ($T_{\rm RH}\ge 10^{13}$~GeV), which gives the most conservative result. The right panel shows a zoom over the orange region, indicating how the bounds change with the measured value of $r$.
}
\label{fig: mt alpha Zoom}
\end{figure}
As shown in fig.~\ref{fig: mt xi 2}, the bound gets weaker as $T_{\rm RH}$ increases since thermal effects are able to rescue the Higgs field, driving it away from the instability region and effectively increasing the value of $h_{\rm end}$.
 The temperature effect saturates around $T_{\rm RH}\approx 10^{11}$~GeV. For larger reheating temperatures, the bound is practically independent of $T_{\rm RH}$ since $h_{\rm end}$ converges to $h_{\rm cl}$.

In fig.~\ref{fig: mt mh} we show how the upper bound from Higgs stability is affected by varying $m_h$. Changing the Higgs mass by $2\sigma$ is equivalent to a change in $m_t$ of about $0.3\sigma$. This means that the present experimental determination of $m_h$ is so precise that we can safely fix the Higgs mass to its central value and neglect the uncertainty associated with $m_h$.

Another consideration is that the orange band in figs.~\ref{fig: mt xi} and~\ref{fig: mt xi 2}, in spite of our generous range of detectability ($5\times 10^{-4}<r<6.4 \times 10^{-2}$), is fairly narrow. This means that the bound does not depend significantly on the actual measurement of $r$, but only on a positive detection. For this reason, in fig.~\ref{fig: mt alpha xi} we fix 
a typical value of the tensor-to-scalar ratio to the current experimental bound ($r= 0.064$)
 and show contour lines of the corresponding upper bound on $\xi$ in the plane $m_t$--$\alpha_s(m_Z)$. We have chosen a large value of $T_{\rm RH}$, which corresponds to the most conservative situation to set upper bounds on $\xi$. Any other assumption on $T_{\rm RH}$ will make the bound on $\xi$ stronger. Figure~\ref{fig: mt alpha xi} shows that, even under the most conservative assumption on $T_{\rm RH}$, a measurement of $r$ will provide a significant constraint on $\xi$. The bound critically depends on $m_t$ and $\alpha_s$, and any experimental improvement on their determination will carry important information. Alternatively, fig.~\ref{fig: mt alpha xi} can be read as a bound on $m_t$ and $\alpha_s$, for a given assumption on the value of $\xi$: a measurement of $r$ would rule out the region to the right of the line corresponding to the chosen value of $\xi$. 
 The lines overlap and become indistinguishable for any value of $\xi$ larger than about 0.02 (but still $\xi\lsim 0.1$). In this range the bound from Higgs stability is fairly independent of the values of $\xi$ and $T_{\rm RH}$ (see fig.~\ref{fig: mt xi 2}) and affects only the SM parameters. 

For positive values of $\xi\lsim 0.1$, the Higgs potential acquires a local minimum driven by the negative contribution to the effective squared mass from the $\xi$-term, but the stochastic fluctuations are still active, in contrast to what we discuss in the next section.
The location of the barrier of the Higgs potential and its value at the maximum are also not significantly affected. As a result, the dynamics does not depend much on the value of $\xi$. 
In fig. \ref{fig: mt alpha Zoom} we show how the regions are distributed for the special value $\xi =0$, as a function of $\alpha_s$ and $m_t$.

\begin{figure}[h!] \centering
\begin{subfigure}[t]{0.49\textwidth}\includegraphics[width=\textwidth]{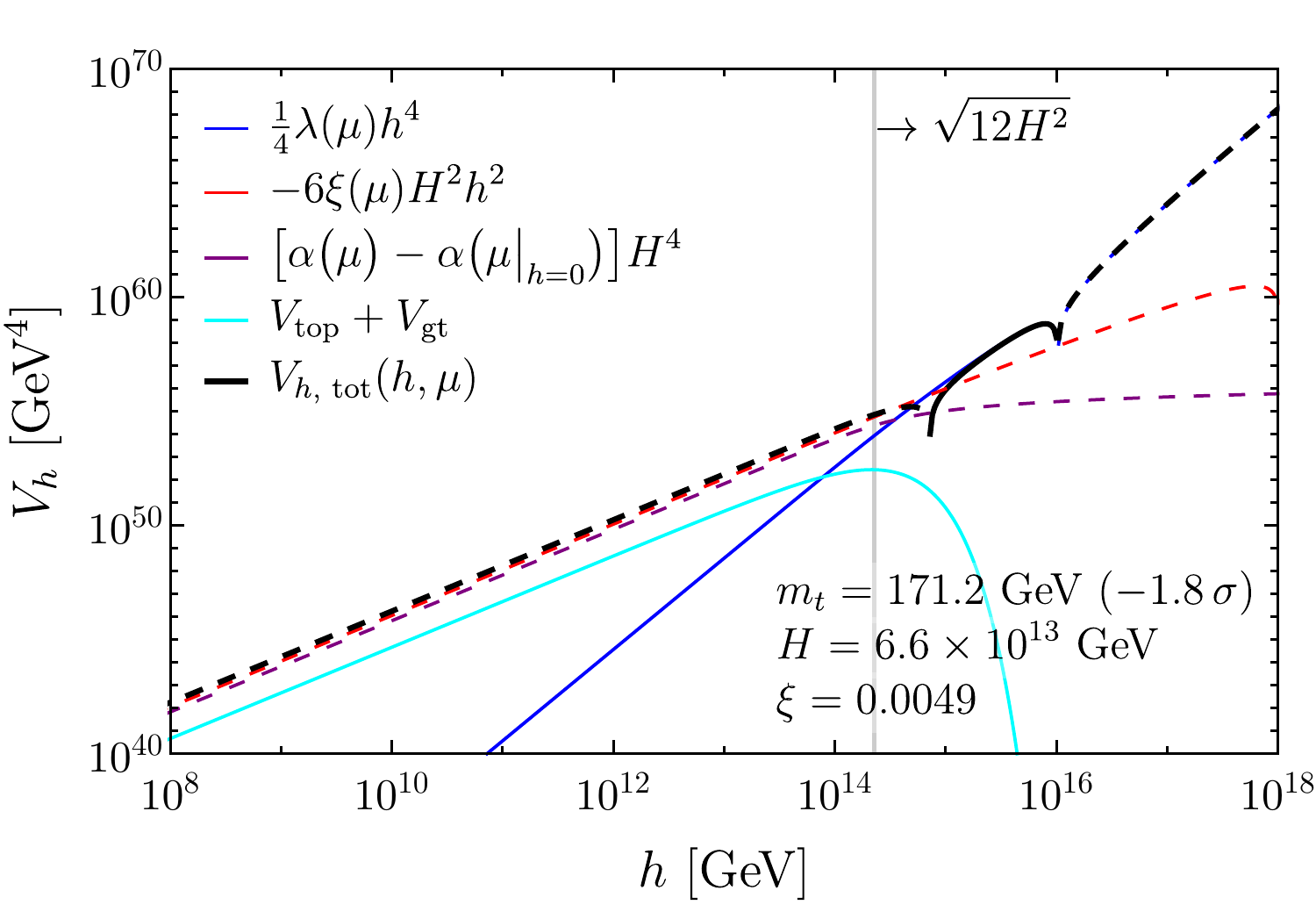} \centering
\caption{\textcolor{verde}{\textit{Green}}}
\label{fig plot Vh 1}
\end{subfigure}
\begin{subfigure}[t]{0.49\textwidth}
\includegraphics[width=\textwidth]{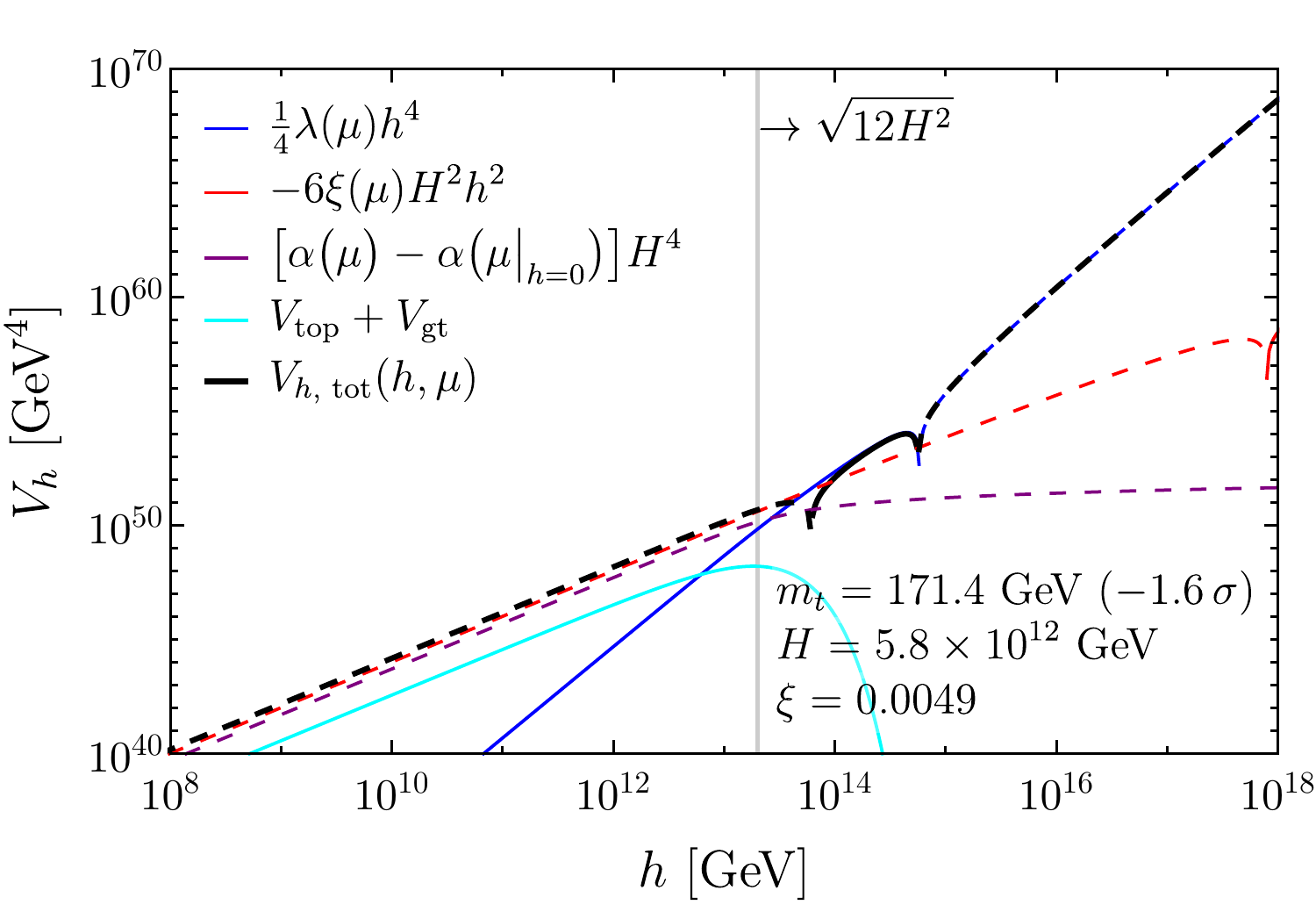} 
\caption{\textcolor{arancione}{\textit{Orange}}}
\label{fig plot Vh 2}
\end{subfigure}
\begin{subfigure}[t]{0.49\textwidth}
\includegraphics[width=\textwidth]{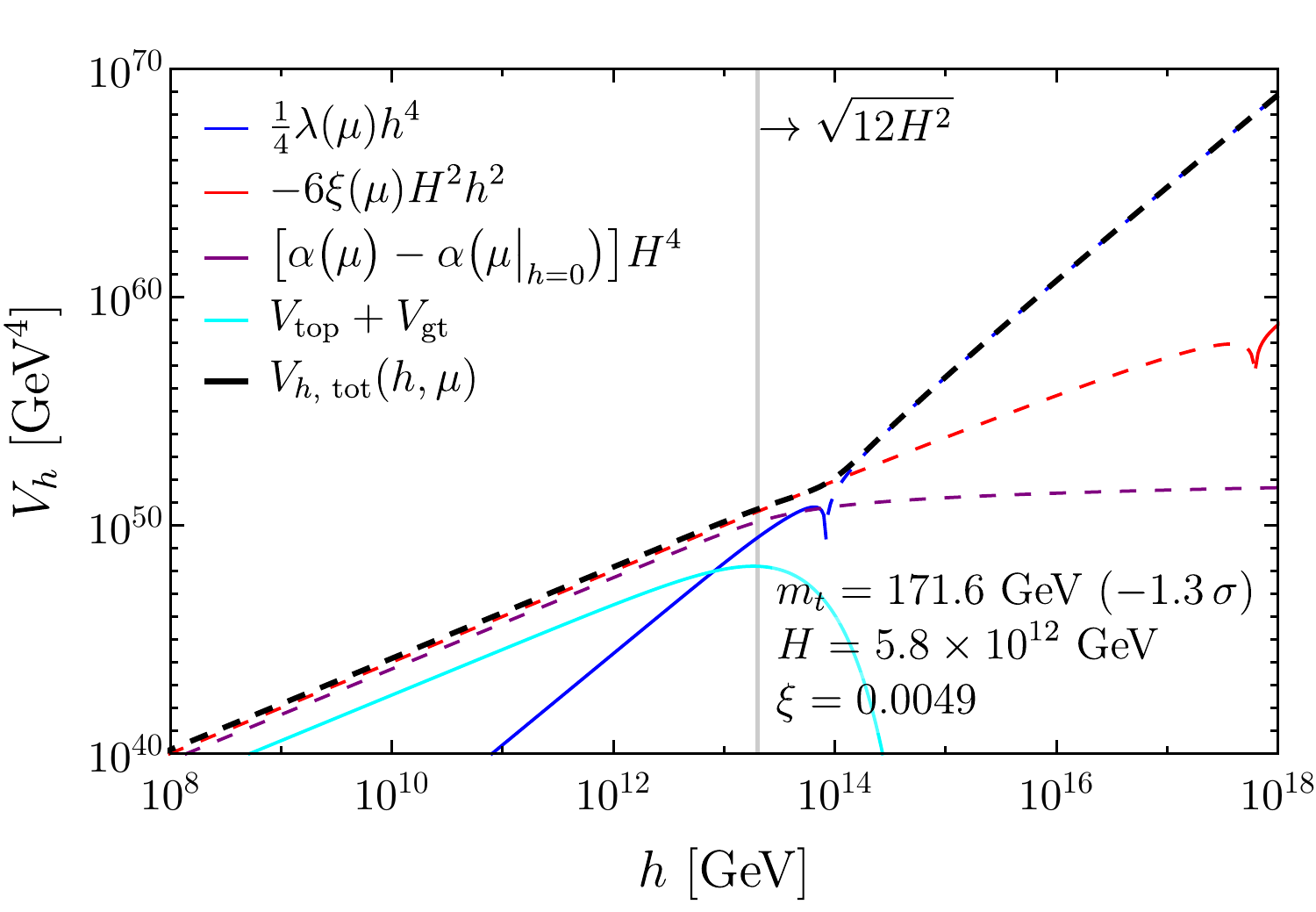}
\caption{\textcolor{rosso}{\textit{Red}}}
\label{fig plot Vh 3}
\end{subfigure}
\begin{subfigure}[t]{0.49\textwidth}
\includegraphics[width=\textwidth]{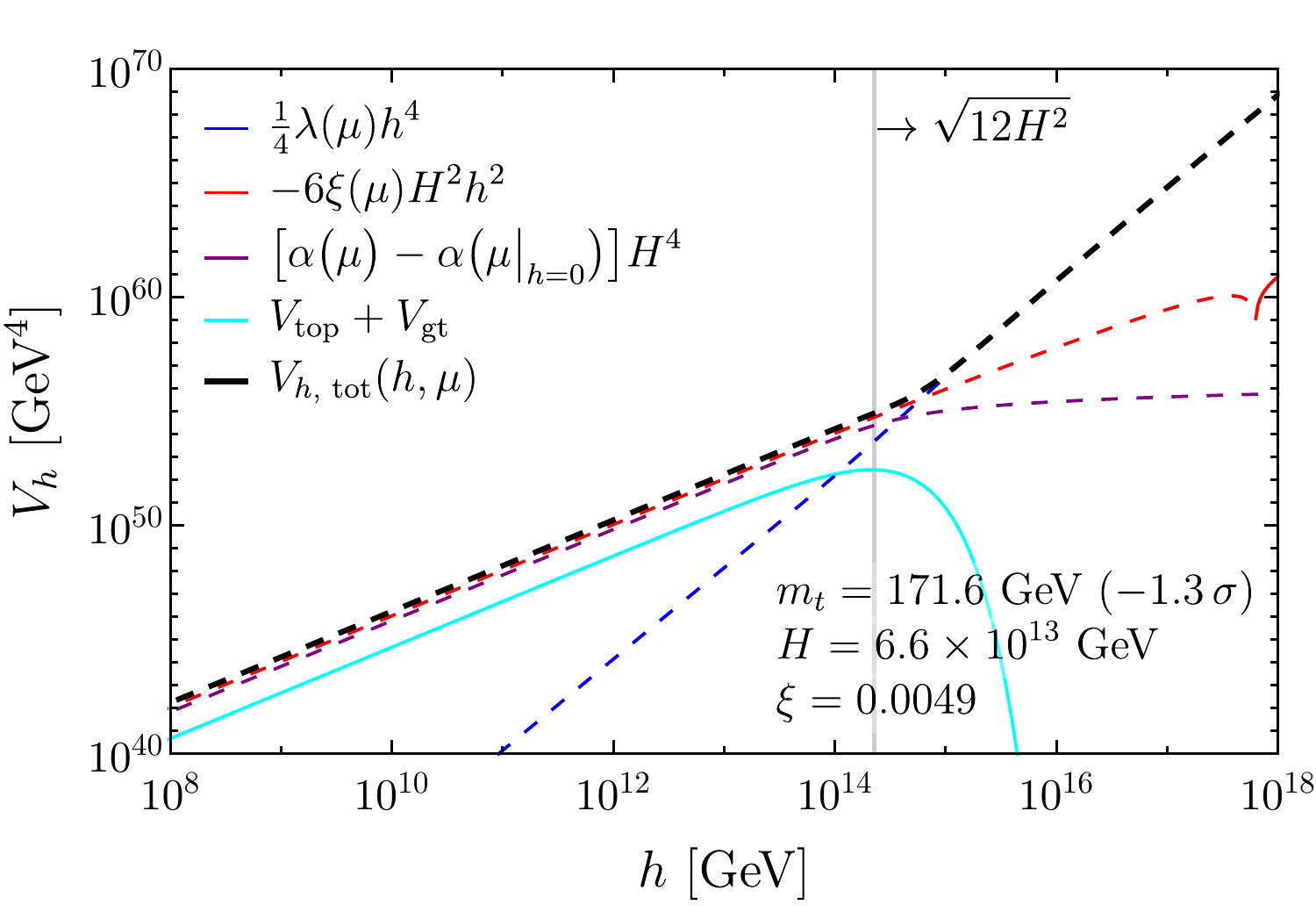}
\caption{\textcolor{rosso}{\textit{Red}}}
\label{fig plot Vh 4}
\end{subfigure}
\caption{The Higgs potential during inflation (black line) and its separate contributions (coloured lines) for different choices of $m_t$ and $H$. We have fixed $m_h$ and $\alpha_s$ to their central values and the Higgs-curvature coupling (evaluated at the scale $6\times 10^{13}$~GeV) to the value $\xi=0.0049$.
    Dashed lines indicate that the corresponding term in the Higgs potential is negative.}
\label{fig: plot Vh}
\end{figure}

%\begin{figure}[h!] \centering
%\includegraphics[width=.45\textwidth]{mt-alpha-Regions}
%\includegraphics[width=.45\textwidth]{mt-alpha-Regions-Zoom}
%\caption{Bounds as in fig.~\ref{fig: mt xi}, valid for $\xi \simgt 3/32$. The bounds are independent of $T_{\rm RH}$ and become stronger as $\xi$ grows, although only with a mild logarithmic dependence. The right panel shows a zoom around the orange region, indicating how the bounds change with the measured value of $r$.
%}
%\label{fig: mt alpha Zoom}
%\end{figure}

\subsection{Case {\boldmath $ \xi\gsim 0.1$} }
\label{secxi01}
As we already mentioned, for $\xi\gsim 0.1$, a new minimum (see fig.~\ref{fig plot Vh 1} and \ref{fig plot Vh 2}) is created such that
the Higgs mass squared around it is large enough to damp the Higgs fluctuations. Therefore, 
the Higgs field during inflation is frozen at the new minimum. At the end of inflation, the Higgs is driven back towards the origin, without causing any concern for electroweak stability. However, if the quartic coupling $\lambda$ becomes negative before the Higgs reaches the new minimum (as in figs.~\ref{fig plot Vh 3} and \ref{fig plot Vh 4}), then the Higgs potential is always negative and the classical evolution drives the Higgs field towards the dangerous AdS region. 

\begin{figure}[h!] \centering
\includegraphics[width=.435\textwidth]{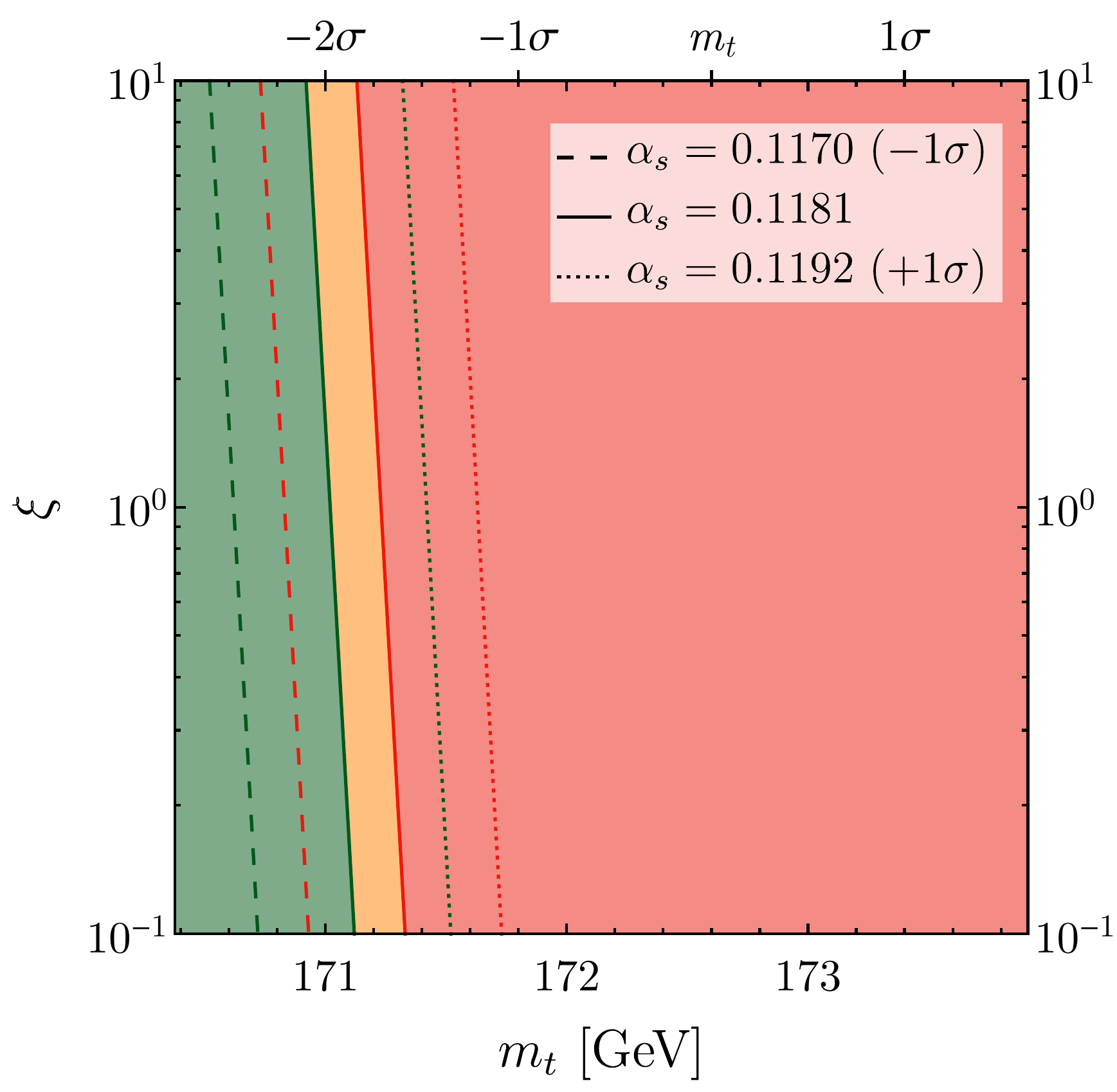}
\caption{
As in fig.~\ref{fig: mt xi}, for $\xi > 0.1$. The results are independent of $T_{\rm RH}$.
}
\label{fig: mt xi xi gtr 0.1}
\end{figure}

\begin{figure}[h!] \centering
\includegraphics[width=.435\textwidth]{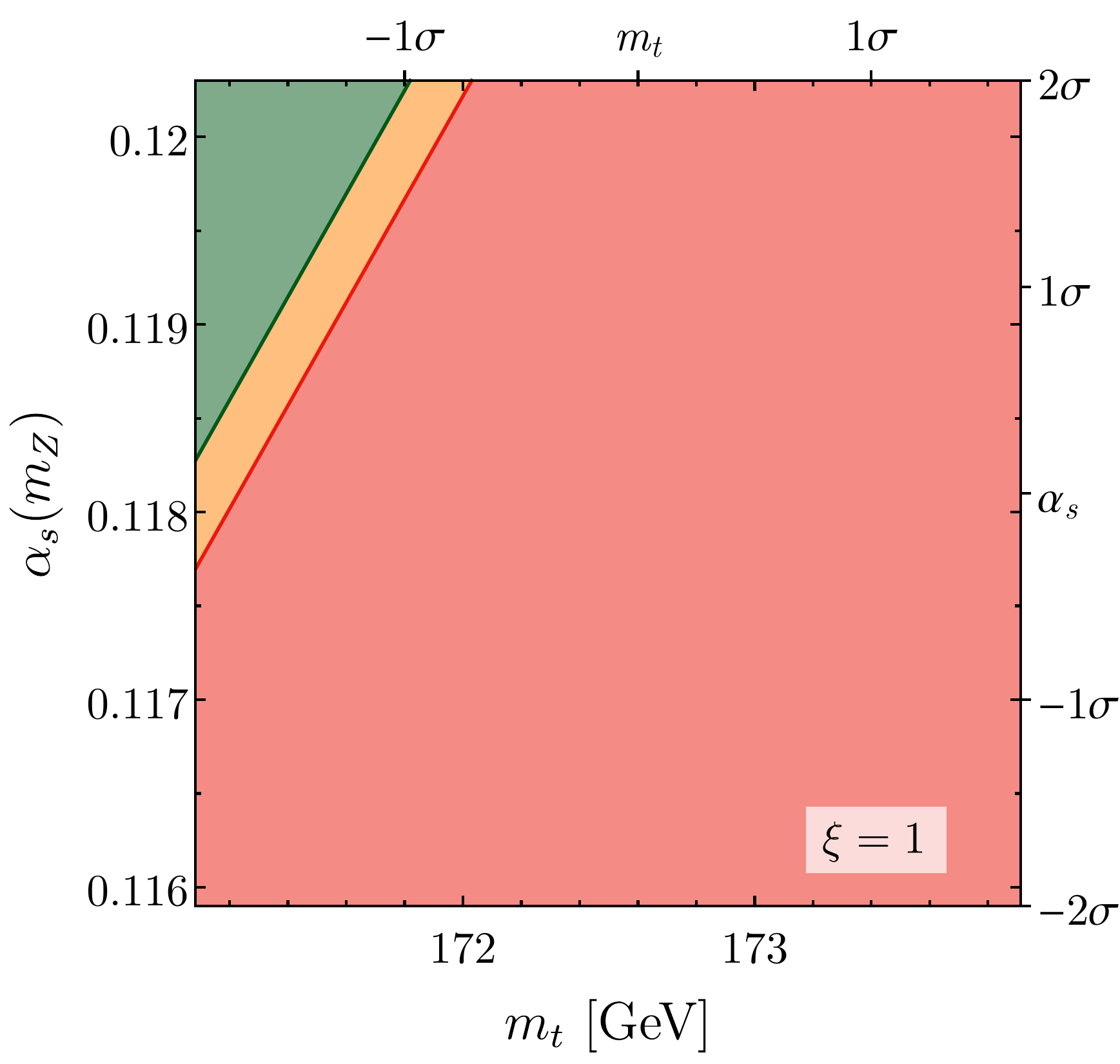}
\includegraphics[width=.435\textwidth]{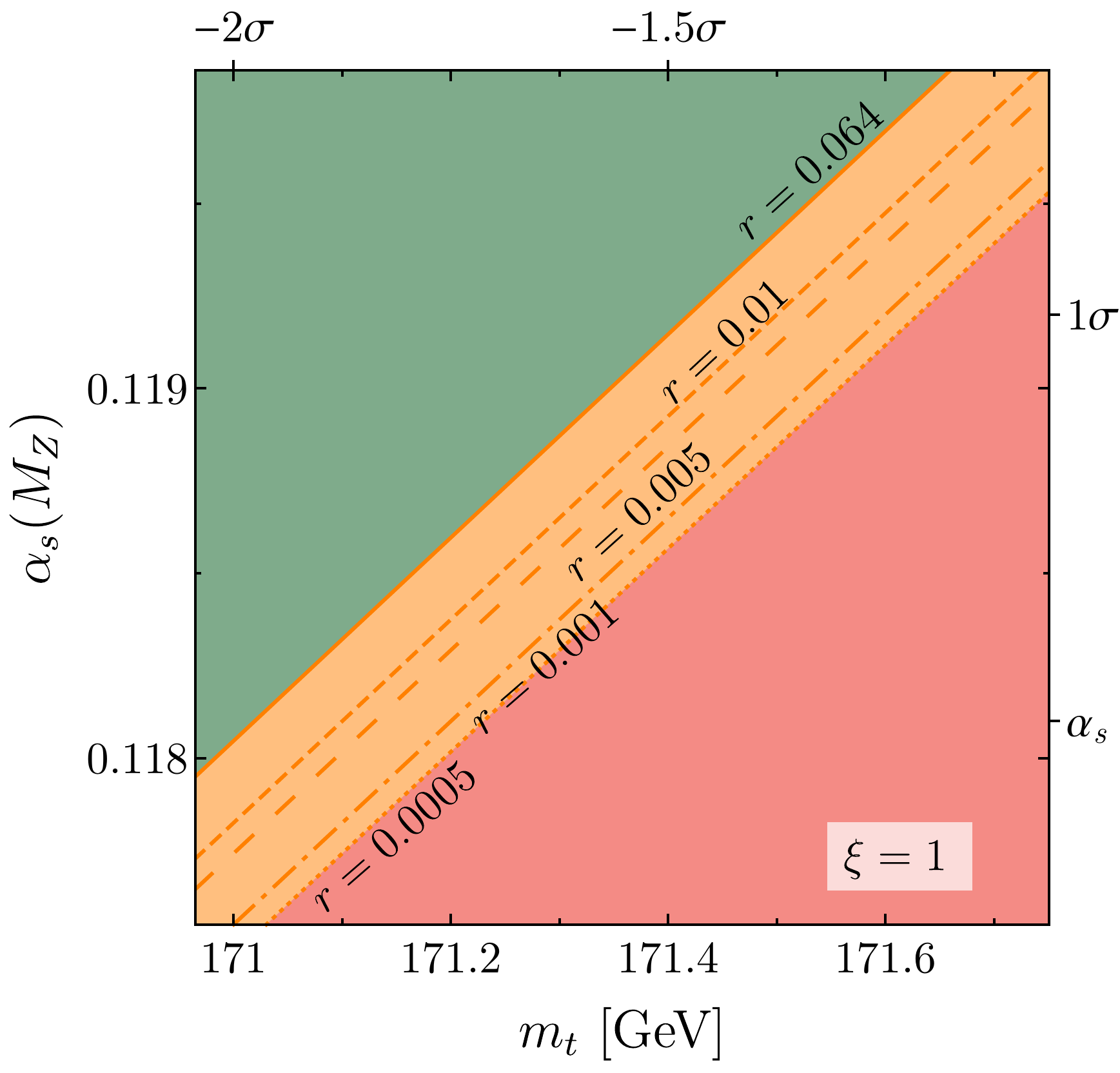}
\caption{
\textit{Left:} As in fig.~\ref{fig: mt alpha Zoom}, for $\xi =1$. The results are independent of $T_{\rm RH}$ and relatively robust against the (logarithmic) variation of $\xi$, valid whenever $\xi \simgt 0.1$.
\textit{Right}:  Zoom over the orange region, indicating how the bounds change with the measured value of $r$.
%The bounds are independent of $T_{\rm RH}$ and become stronger as $\xi$ grows, although only with a mild logarithmic dependence. 
}
\label{fig: xi gtr 0.1}
\end{figure}

Therefore, for $\xi \simgt 0.1$, the Higgs evolution is essentially classical and the stability condition
is obtained simply by imposing that a minimum exists. In fig.~\ref{fig: mt xi xi gtr 0.1} we plot, for $\xi \simgt 0.1$, the bound coming from a hypothetical measurement of $r$ on the plane $(m_t,\xi)$. Given the weak sensitivity on $\xi$, in fig.~\ref{fig: xi gtr 0.1} we fix $\xi =1$ and show the bound on $m_t$ and $\alpha_s(m_Z)$. The result is approximately valid within a wide range of $\xi$, as long as $\xi \simgt 0.1$.
A good numerical  fit parametrising the behaviour of figs.~\ref{fig: mt xi xi gtr 0.1} and~\ref{fig: xi gtr 0.1} is given by 
\be
m_t \simlt 171.2~{\rm GeV} +0.4  \left( \frac{\alpha_s - 0.1181}{0.0011} \right) - 0.1\left[ \log_{10}\left( \frac{r}{10^{-2}}\right)+
\log_{10}\left(\frac{\xi}{10^{-1}}\right)\right] \, .
\label{estim99}
\ee
The bound is quite significant and can rule out a vast portion of SM parameters. It is independent of $T_{\rm RH}$ and it becomes (logarithmically) stronger with $\xi$.

The parametric behaviour of eq.~(\ref{estim99}) can be understood through a simple estimate. The Higgs stability condition corresponds to the requirement that 
 the new local  minimum at $\langle h \rangle \simeq H \sqrt{12\, \xi /\lambda}$ is developed before the instability scale is reached.
 Identifying, for simplicity, the instability scale with $\Lambda$, the Higgs stability condition becomes $H \sqrt{12\, \xi /\lambda}\simlt \Lambda$. Using a simple analytical estimate of the scale $\Lambda$ 
 \be
\log_{10}\left(\frac{\Lambda}{\rm GeV} \right)\simeq 12 - 1.4 \left( \frac{m_t/{\rm GeV} -172.6}{0.8}\right) +0.6 \left( \frac{\alpha_s - 0.1181}{0.0011} \right) \, ,
\ee
we reproduce the parametric behaviour in eq.~(\ref{estim99}).

%The bounds are independent of $T_{\rm RH}$ and become stronger as $\xi$ grows, although only with a mild logarithmic dependence.
For large positive values of $\xi$ the green region extends vertically without significant changes, following the mild logarithmic dependence. However, for values of $\xi$ larger than $\sim  10^7$, the potential becomes always unstable when $H\simgt 6 \times 10^{12}$ GeV since the instability is given by the dominant negative quadratic term in the potential up to (at least) the Planck scale. 
%\DRcanc{The green regions in figs.~\ref{fig: mt xi} and~\ref{fig: mt xi 2}  extend to  low values of $m_t$ for $\xi< 10^7$ (at the Planck scale) since, for sufficiently small $m_t$, the quartic $\lambda(\mu)$ is always positive up to very large energies.}
These considerations are independent of the reheating temperature.

%
%This estimate shows that the dependence on $r$ and $\xi$ is only logarithmic. \GG{Il mio valore di 170 GeV e' un po' bassino rispetto alla figura. Anzi, se prendo i numeri mi viene 169 GeV. Se usassi 11, invece di 10, per la scala di instabilita, troverei 171 GeV, che e' meglio. Vi torna numericamente la dipendenza logaritmica in $\xi$?}

Finally, we have checked that all our results are robust against modifying the number of $e$-folds ({\it e.g.} taking $N=50$) or changing the normalisation scale \eqref{n} ({\it e.g.} taking $\mu\approx \sqrt{h^2+H^2}$).
Note that the  choice of the renormalisation scale in eq.~\eqref{n} may considerably change the shape of the potential with respect to the choice $\mu \simeq h$, since   the running of the  quartic coupling $\lambda(\mu)$ and of $\xi(\mu)$ are frozen for  $h\ll H$. 
One can see this effect by comparing figs.~\ref{fig plot Vh 3} and \ref{fig plot Vh 4}, where the only difference in the parameters is the choice of the Hubble rate.
Furthermore, fig.~\ref{fig: plot Vh} shows that the contributions to the effective potential from infrared tops and gauge bosons produced during inflation is much smaller than the quadratic piece proportional to the non-minimal $\xi$-coupling to gravity.

%%%%%%%%%%%%%%%%%%%%%%%%%%%%

\section{Conclusions} 
\noindent
The detection of inflationary tensor modes in future experiments measuring the B-mode polarisation would have paramount implications for our understanding of the early history of the universe. It would act as a powerful discriminator for inflationary models and establish the occurrence of super-Planckian field excursions during inflation~\cite{Lyth:1996im}. It would exclude most models where the Peccei-Quinn symmetry is broken during inflation, because of unacceptable axion isocurvature fluctuations~\cite{Marsh:2014qoa,Visinelli:2014twa}. It would also quantify the relation between the reheating temperature after inflation (defined in terms of the inflaton decay width $\Gamma_\phi$) and $T_{\rm max}$, the maximum temperature attained by the thermal bath originating from the inflaton decay process:
\beq
\frac{T_{\rm max}^2}{T_{\rm RH}} = \left( \frac{r}{10^{-2}}\right)^{1/4} \, 1.6 \times 10^{15}~{\rm GeV} \, 
\eeq
(where the numerical coefficient assumes SM degrees of freedom in the thermal bath)\footnote{As a side remark, note that finding $T_{\rm max}\gg T_{\rm RH} $ would not impact the bounds on dangerous thermal relics. Even after properly taking into account the detailed reheating dynamics~\cite{Giudice:2000ex}, the limits on gravitino thermal production~\cite{Giudice:1999am} or on the viability of leptogenesis~\cite{Giudice:2003jh} depend only on $T_{\rm RH}$, and not on $T_{\rm max}$. The reason is that, during the initial reheating phase when the temperature can be larger than $T_{\rm RH}$, there is a large injection of entropy that dilutes the relative abundance of new particles. In other words, the effective energy density available during reheating is  $\Gamma_\phi^2 M_{\rm Pl}^2$, and not $H^2 M_{\rm Pl}^2$.}.

In this paper we have discussed another interesting consequence of the measurement of primordial tensor modes: the constraints on SM parameters coming from the requirement that inflation does not destabilise the electroweak vacuum, under the assumption that the SM is valid up to the instability scale. The basic observation is that a measure of the tensor-to-scalar ratio will pin down the value of the energy scale during primordial inflation, which is the crucial parameter in assessing if Higgs stochastic fluctuations are
lethal in the early stages of the evolution of the universe. We have used state-of-the-art results as ingredients of our calculation: the Higgs potential is evaluated at NNLO precision~\cite{buttazzo}; one-loop curvature correction in de Sitter space are included and an optimal choice for the renormalisation scale is made, see eq.~(\ref{n})~\cite{scale}; we include the effect of gravitational production of top quarks~\cite{f} and gauge bosons during inflation (although the effect is quantitatively negligible); we follow numerically the evolution of the Higgs field during and after inflation, thus making our results robust against any particular choice for analytic criteria defining ``conditions for safeness". 

Our results can be summarised as follows. Under the assumption that the SM can be extrapolated up to very high energies, a detection of the tensor-to-scalar ratio $r$ implies a bound on SM parameters. Because of the relatively narrow window of opportunity for detection ($5\times 10^{-4} \lsim r \lsim 6\times 10^{-2}$), the bound on SM parameters does not depend much on the measured value of $r$. The bound is also fairly insensitive to the Higgs mass, given the high precision with which this parameter is known today. Thus, the bound involves the following four parameters: the top mass $m_t$, the strong coupling $\alpha_s$, the Higgs-curvature coupling $\xi$, and the reheating temperature $T_{\rm RH}$. Schematically, the bound can be characterised by three cases, which correspond to different physical situations.

\begin{itemize}

\item {\bf Case} {\boldmath $ \xi \lsim -3/16.$} No bound is obtained, since Higgs fluctuations during inflation are efficiently dumped.

\item {\bf Case} {\boldmath $ -3/16 \lsim \xi \lsim 0.1.$} The bound on $m_t$ and $\alpha_s$ gets progressively weaker as $T_{\rm RH}$ is raised, as shown in figs.~\ref{fig: mt xi}--\ref{fig: mt xi 2}. For $T_{\rm RH}\gsim 10^{11}$~GeV, the bound saturates and becomes independent of $T_{\rm RH}$. In this case, the results are shown in fig.~\ref{fig: mt alpha xi}, which can be viewed as an upper bound on $\xi$, for given $m_t$ and $\alpha_s$, or as a bound on the SM parameters, for any given $\xi$. The bounds shown in fig.~\ref{fig: mt alpha xi} are valid for any $T_{\rm RH}$, and only become stronger by assuming a low reheating temperature.

\item {\bf Case} {\boldmath $\xi \gsim 0.1.$} The Higgs dynamics is dominated by classical evolution. Either the Higgs is frozen during inflation at a new minimum and protected from quantum fluctuations, or is driven classically to the dangerous AdS region. The corresponding condition discriminating between these two cases and establishing Higgs stability is shown
 in figs.~\ref{fig: mt xi xi gtr 0.1} and~\ref{fig: xi gtr 0.1}. The bound on SM parameters is insensitive to $T_{\rm RH}$ and has only a logarithmic dependence on $\xi$. The resulting constraint on $m_t$ and $\alpha_s$ (shown in fig.~\ref{fig: xi gtr 0.1}) is quite stringent and the survival of the electroweak vacuum can be obtained only for fairly low values of $m_t$ and/or high values of $\alpha_s$.

\end{itemize}

\subsection*{Acknowledgments}
We thank D. Rodriguez-Roman and  M. Fairbairn for useful correspondence about ref.~\cite{f}. A.~R.\ and D.~R.\ are supported by the Swiss National Science Foundation (SNSF),  project {\sl The Non-Gaussian Universe and Cosmological Symmetries}, project number: 200020-178787.  
This research was supported in part by Perimeter Institute for Theoretical Physics. Research at Perimeter Institute is supported by the Government of Canada through the Department of Innovation, Science and Economic Development and by the Province of Ontario through the Ministry of Research, Innovation and Science.

\end{document}